\documentclass[a4paper,11pt]{article}
\pdfoutput=1 

\usepackage{jheppub} 
\usepackage{neuralnetwork}
\usepackage{caption}
\usepackage{subcaption}
\usepackage{multicol}
\usepackage{graphicx} 
\usepackage{cancel}
\usepackage{grffile}
\usepackage{braket}

\usepackage[T1]{fontenc} 

\author[a, b, c]{Melissa van Beekveld}
\author[b,c]{Sascha Caron}
\author[b]{Luc Hendriks}
\author[d]{Paul Jackson}
\author[d]{Adam Leinweber}
\author[b,e]{Sydney Otten}
\author[d]{Riley Patrick}
\author[f]{Roberto Ruiz de Austri}
\author[d]{Marco Santoni}
\author[d]{and Martin White}

\affiliation[a]{Rudolf Peierls Centre for Theoretical Physics, Clarendon Laboratory, Parks Road, Oxford OX1
3PU, UK}
\affiliation[b]{High Energy Physics, IMAPP, Radboud University Nijmegen, Heyendaalseweg 135, 6525 AJ
Nijmegen, NL}
\affiliation[c]{Nikhef, Science Park 105, 1098 XG Amsterdam, NL}
\affiliation[d]{ARC Centre of Excellence for Dark Matter Particle Physics, University of Adelaide, North Terrace, SA 5005}
\affiliation[e]{Gravitation and Astroparticle Physics Amsterdam (GRAPPA), Science Park 904, 1098 XH Amsterdam, NL}
\affiliation[f]{Instituto de F\'isica Corpuscular, IFIC-UV/CSIC, Valencia, Spain}

\emailAdd{melissa.vanbeekveld@physics.ox.ac.uk}
\emailAdd{luchendriks@gmail.com}
\emailAdd{adam.leinweber@adelaide.edu.au}

\title{\boldmath Combining outlier analysis algorithms to identify new physics at the LHC}

\abstract{
The lack of evidence for new physics at the Large Hadron Collider so far has prompted the development of model-independent search techniques. In this study, we compare the anomaly scores of a variety of anomaly detection techniques: an isolation forest, a Gaussian mixture model, a static autoencoder, and a $\beta$-variational autoencoder (VAE), where we define the reconstruction loss of the latter as a weighted combination of regression and classification terms. We apply these algorithms to the 4-vectors of simulated LHC data, but also investigate the performance when the non-VAE algorithms are applied to the latent space variables created by the VAE. In addition, we assess the performance when the anomaly scores of these algorithms are combined in various ways. Using supersymmetric benchmark points, we find that the logical AND combination of the anomaly scores yielded from algorithms trained in the latent space of the VAE is the most effective discriminator of all methods tested.
}

\begin{document} 
\maketitle
\flushbottom

\section{Introduction}
\label{sec:intro}
The 2012 discovery of the Higgs boson puts us at an intriguing milestone in the history of particle physics: the next discovery in collider physics is likely to arise from a theory whose precise details will not be known in advance. Whilst a plethora of well-motivated extensions to the Standard Model (SM) of particle physics have been proposed, there is no unambiguous prediction of the phenomenology that we can expect to observe at the Large Hadron Collider (LHC). Many searches for theories of interest such as supersymmetry are currently heavily optimised on specific benchmark scenarios, an approach which kills sensitivity to a significant bulk of viable SUSY models~\cite{Aad:2015baa,Khachatryan:2016nvf,Athron:2017qdc,Athron:2017yua,Athron:2018vxy}.

In recent years, a number of techniques have been developed for performing signal model-independent searches with collider data. The D0 collaboration at the Tevatron developed an unsupervised, multivariate signal detection algorithm named SLEUTH~\cite{Abbott:2000fb,Abbott:2000gx,Abbott:2001ke, Abazov:2011ma}, the H1 Collaboration~\cite{Aktas:2004pz,Aaron:2008aa} at HERA used a 1-dimensional  signal detection algorithm, 
and the CDF Collaboration~\cite{Aaltonen:2007dg, Aaltonen:2008vt} at the Tevatron also developed a 1-dimensional signal-detection algorithm. The BUMPHUNTER algorithm continues these efforts at the LHC~\cite{Choudalakis:2011qn}. Other model-independent LHC searches have been performed by the ATLAS and CMS collaborations, or with their publicly available data~\cite{Aaboud:2018ufy,CMS:2008gya,ATLAS:2017irs,Asadi:2017qon,CMS:2011fra}. A promising approach that uses neural networks to compare observations with a set of reference events (via the definition of a suitable test statistic) was presented in~\cite{DAgnolo:2018cun}. The use of autoencoders in jet substructure applications has been outlined in~\cite{Farina:2018fyg,Heimel:2018mkt}, whilst the use of autoencoders for general LHC searches was also explored in~\cite{Hajer:2018kqm}, with a particular emphasis on proposing new anomaly score metrics that can increase the likelihood that anomalous data will be identified. Variational autoencoders have been used to detect anomalies using high level features as inputs in~\cite{cerri2019variational}, and using adversarial neural networks in \cite{knapp2020adversarially}.  An early use of Gaussian mixture models to find anomalies in the context of Higgs boson searches is detailed in CDF~\cite{Kuusela:2011aa}. Further model-independent or weakly-supervised techniques for LHC discovery applications have been proposed in~\cite{Andreassen:2020nkr,Nachman:2020lpy,Collins:2019jip,Dery:2018dqr,Collins:2018epr,Benkendorfer:2020gek}.

In this paper, we perform a systematic comparison of a variety of anomaly detection techniques for LHC searches. Throughout, our aim is to define an anomaly-score variable that can be used on an \textit{event-by-event} basis to classify a given event as a potential signal of new physics (yielding score $\approx 1$) or stemming from the SM (yielding a score $\approx 0$). The underlying assumption is that some signals of new physics are different in the kinematics (defined by 4-vectors) and object types from SM events.
We first determine the extent to which a variety of unsupervised anomaly-detection techniques  can detect anomalies when used on final state multiplicities and 4-vector information at the LHC. The considered techniques are an isolation forest (IF), a Gaussian mixture model (GMM), a static autoencoder (AE), and a variational autoencoder (VAE) architecture where we define the reconstruction loss as a weigthed combination of regression and classification terms. We use a selection of supersymmetric benchmark models to assess the performance on the hypothetical signals, without optimising the hyperparameters of our unsupervised techniques. We then assess various ``combination'' techniques that explore ways to combine the results of each anomaly detection method, such as taking a logical OR or AND of the different anomaly  scores, or taking the product/average of the scores. We then investigate how the performance changes when the non-VAE techniques are run on the latent space variables of the VAE, supplemented by the VAE reconstruction loss. We perform the same combination techniques as in the earlier case, and compare the performance of combinations of algorithms trained on 4-vectors to algorithms trained on latent space representations. Taking one particular combination of the loss parameters, we find strong performance on both gluino pair-production and stop pair-production scenarios. 

This paper is structured as follows. In Section \ref{sec:example}, we provide the details of our SM data and the supersymmetric benchmark models. In Section~\ref{sec:method} we define and describe the VAE, isolation forest, Gaussian mixture model and static autoencoder algorithms. Results are presented in Section~\ref{sec:results} and a short summary is presented in Section~\ref{sec:summary}, before we present our conclusions in Section~\ref{sec:conclusion}.

\section{Selection of processes and generation of events}
\label{sec:example}
Although our proposed techniques are designed to provide a model-independent approach to LHC searches, it is useful to test their performance on particular models of interest. To this end, we use a variety of supersymmetric benchmark models to test our techniques, using the supersymmetric signal and SM background processes from the dataset published and described in Ref.~\cite{Brooijmans:2020yij}.

The events are generated at leading order for a $13$ TeV LHC centre of mass energy using \texttt{Madgraph v6.3.2}~\cite{Alwall:2014hca} with the NNPDF PDF set~\cite{Buckley:2014ana} in the $5$ flavour proton scheme. {\tt Madgraph} was interfaced with \texttt{Pythia 8.2}~\cite{Sjostrand:2014zea} for the parton shower (using MLM matching) and detector effects are included using \texttt{Delphes 3}~\cite{deFavereau:2013fsa} with a modified version of the ATLAS detector card. A summary of the supersymmetric benchmark models and SM backgrounds used can be seen in Tab~\ref{tab:bsm-proc} and \ref{tab:sm-proc} respectively.

Our algorithms are developed and trained on the SM background processes only, which will in general give worse performance than a supervised approach that assumes knowledge of the signal. However, it is important to stress that the question we wish to answer is ``to what extent would our techniques be able to discover or exclude interesting models if one does not know about them in advance?''.

\begin{table}[!t]
\centering
\begin{tabular}{|l|l|r|r|}
\hline
Process & Process ID & $\sigma$ (pb) & $N_{\text{tot}}$ ($N_{10 \text{fb}^{-1}}$) \\\hline\hline
$pp \to \tilde{g}\tilde{g}$ ($1$ TeV)     & Gluino 01 & 0.20 & 50000 (2013)\\
$pp \to \tilde{g}\tilde{g}$ ($1.2$ TeV)	  & Gluino 02 &0.05 & 50000 (508)\\
$pp \to \tilde{g}\tilde{g}$ ($1.4$ TeV)	  & Gluino 03 &0.014 & 50000 (144)\\
$pp \to \tilde{g}\tilde{g}$ ($1.6$ TeV)	  & Gluino 04 &0.004 & 50000 (44)\\
$pp \to \tilde{g}\tilde{g}$ ($1.8$ TeV)	  & Gluino 05 &0.001 & 50000 (14)\\
$pp \to \tilde{g}\tilde{g}$ ($2$ TeV)	  & Gluino 06 &$4.8\times10^{-4}$ & 50000 (5)\\
$pp \to \tilde{g}\tilde{g}$ ($2.2$ TeV)	  & Gluino 07 &$1.7\times10^{-4}$ & 50000 (2)\\\hline
$pp \to \tilde{t}_1\tilde{t}_1$ (220 GeV), $m_{\tilde{\chi}^0_1} = 20$ GeV 		& Stop 01 & 26.7 & 500000 (267494) \\
$pp \to \tilde{t}_1\tilde{t}_1$ (300 GeV), $m_{\tilde{\chi}^0_1} = 100$ GeV  	& Stop 02 & 5.7 & 500000 (56977)   \\
$pp \to \tilde{t}_1\tilde{t}_1$ (400 GeV), $m_{\tilde{\chi}^0_1} = 100$ GeV 		& Stop 03 & 1.25 & 250000 (12483)  \\
$pp \to \tilde{t}_1\tilde{t}_1$ (800 GeV), $m_{\tilde{\chi}^0_1} = 100$ GeV  	& Stop 04 & 0.02 & 250000 (201)    \\
\hline
\end{tabular}
\caption{\label{tab:bsm-proc} Summary of the supersymmetric benchmark models that are used to test our methods. The details include the production cross-section at $\sqrt{s}=13$ TeV, the number of events that were generated, and the number of events expected in $10 \text{fb}^{-1}$ of LHC data ~\cite{Brooijmans:2020yij}.}
\end{table}

\begin{table}[!t]
\centering
\begin{tabular}{|l|l|r|r|}
\hline
Process & Process ID &  $\sigma$ (pb) & $N_{\text{tot}}$ ($N_{10 \text{fb}^{-1}}$) \\\hline\hline
$pp \to jj$ 					  & njets & 19718$_{H_T>600 \text{GeV}}$ & 415331302 (197179140)\\
$pp \to W^{\pm}(+2j) $ 			  & w\_jets & 10537$_{H_T>600 \text{GeV}}$ & 135692164 (105366237)\\
$pp \to \gamma(+2j)$ 			  & gam\_jets & 7927$_{H_T>600 \text{GeV}}$ & 123709226 (79268824)\\
$pp \to Z(+2j)$ 				  & z\_jets & 3753$_{H_T>600 \text{GeV}}$ & 60076409 (37529592)\\
$pp \to t\bar{t}(+2j)$ 			  & ttbar & 541 & 13590811 (5412187)\\
$pp \to W^{\pm}t(+2j)$ 			  & wtop & 318 & 5252172 (3176886)\\
$pp \to W^{\pm}\bar{t}(+2j)$ 	  & wtopbar & 318 & 4723206 (3173834)\\
$pp \to W^+W^-(+2j)$ 			  & ww & 244 & 17740278 (2441354)\\
$pp \to t+\text{jets}(+2j)$ 	  & single\_top & 130 & 7223883 (1297142)\\
$pp \to \bar{t}+\text{jets}(+2j)$ & single\_topbar & 112 & 7179922 (1116396)\\
$pp \to \gamma\gamma(+2j)$ 		  & 2gam & 47.1 & 17464818 (470656)\\
$pp \to W^{\pm}\gamma(+2j)$ 	  & Wgam & 45.1 & 18633683 (450672)\\
$pp \to ZW^{\pm}(+2j)$ 			  & zw & 31.6 & 13847321 (315781)\\
$pp \to Z\gamma(+2j)$ 			  & Zgam & 29.9 & 15909980 (299439)\\ 
$pp \to ZZ(+2j)$ 				  & zz & 9.91 & 7118820 (99092)\\
$pp \to h(+2j)$ 				  & single\_higgs & 1.94 & 2596158 (19383)\\
$pp \to t\bar{t}\gamma(+2j)$ 	  & ttbarGam & 1.55 & 95217 (15471)\\
$pp \to t\bar{t}Z$ 				  & ttbarZ & 0.59 & 300000 (5874)\\
$pp \to t\bar{t}h(+1j)$ 		  & ttbarHiggs & 0.46 & 200476 (4568)\\
$pp \to \gamma t(+2j)$ 			  & atop & 0.39 & 2776166 (3947)\\
$pp \to t\bar{t}W^\pm$ 			  & ttbarW & 0.35 & 279365 (3495)\\
$pp \to \gamma\bar{t}(+2j)$ 	  & atopbar & 0.27 & 4770857 (2707)\\
$pp \to Zt(+2j)$ 				  & ztop & 0.26 & 3213475 (2554)\\
$pp \to Z\bar{t}(+2j)$ 			  & ztopbar & 0.15 & 2741276 (1524)\\
$pp \to t\bar{t}t\bar{t}$ 	   	  & 4top & 0.0097& 399999 (96)\\
$pp \to t\bar{t}W^+W^-$ 		  & ttbarWW & 0.0085& 150000 (85)\\
\hline

\end{tabular}
\caption{\label{tab:sm-proc} Summary of the background processes included in the analysis. The details include the production cross-section at $\sqrt{s}=13$ TeV, the number of events that were generated, and the number of events expected in $10 \, \text{fb}^{-1}$ of LHC data ~\cite{Brooijmans:2020yij}.}
\end{table}

Our first set of BSM models involve supersymmetric gluino pair production, with each gluino subsequently decaying to a boosted top-quark pair and the lightest neutralino, which is stable by assuming $R$-parity conservation. The gluinos are assumed to have a mass of 1-2.2 TeV (in steps of 200 GeV), while the neutralinos have a mass of 1 GeV. The branching ratio of the decay process $\tilde{g}\rightarrow t\bar{t}\tilde{\chi}_1^0$ is taken to be $100\%$. 

In the second scenario two stop quarks ($\tilde{t}_1$) are produced, with each stop decaying into an on-shell top quark and a lightest neutralino ($\tilde{t}_1\rightarrow t\tilde{\chi}^0_1$). We have chosen to take four different benchmark scenarios. In the first model, the lightest neutralino has a mass of 20 GeV and the lightest stop has a mass of 220 GeV. In the other models, the mass of the lightest neutralino is 100 GeV and the stops have masses of 300, 400 and 800 GeV. 

Although the production cross-section for the lowest-mass stop quark pair production is the highest out of all assumed signal hypothesis, it is actually the most difficult to discover in traditional search methods. The mass difference of $\tilde{t}_1$ and $\tilde{\chi}^0_1$ is close to the mass of the top quark, which makes the production of top quark pairs an important irreducible background. The techniques described in the next section are designed to find anomalies, but this model does not result in an obviously anomalous signal. Therefore, we expect that the techniques will show least sensitivity to the $200$~GeV stop scenario, although this might be compensated by the fact that its cross section is the highest. On the other hand, the gluino signals are more anomalous as they result in four top quarks and a sizable missing transverse energy. This is a rare final state for SM production, and since the $1$~TeV gluino carries the highest production cross-section, we expect that this scenario will be the easiest. 

All data is first zero-padded so every event has the same dimensionality. Next, the continuous data and the categorical data are split and the number of objects in the events are counted. From this, the following event structure is defined:

\begin{equation}
\label{eq:event_structure}
\textbf{x} = \left(
N,
\begin{bmatrix}
c_0 \\
c_1 \\
\vdots \\
c_{19}
\end{bmatrix},
\begin{bmatrix}
(p_T, \eta, \phi)_0 \\
(p_T, \eta, \phi)_1 \\
\vdots \\
(p_T, \eta, \phi)_{19}
\end{bmatrix}
\right).
\end{equation}

In this vector, $N$ is the number of objects in the event, $c_i$ is the object type as a one-hot encoded vector, $p_T$ is the transverse momentum, $\eta$ the pseudorapidity and $\phi$ the azimuthal angle of an object. This layout is used to train the unsupervised machine learning algorithms, detailed in Section \ref{sec:method}, on the 4-vector representations of the data. When we later use the non-VAE techniques on the latent space variables of the VAE, it is still true to say that the starting point for the analysis is this 4-vector representation. 

\section{Search methods}
\label{sec:method}
Searches for new physics at the LHC can be divided into two main categories:

\begin{enumerate}
\item {\bf Searches for visibly decaying new particles}, in which all decay products of a new particle are expected to be observed. In this case, one can often find the new physics by observing the invariant mass of the anticipated decay products of the new physics signal, although this becomes problematic in the case that the decay products themselves are unstable (e.g. new resonances decaying to top quarks or gauge bosons), in cases where the width of the resonance is expected to be large, or in cases where strong interference effects distort the shape of the invariant mass peak. 
\item {\bf Searches for semi-invisibly decaying new particles}, in which one cannot rely on the invariant mass to highlight the new physics, and must instead construct various functions of the final state four-vectors in events to try and discriminate the signal from the SM backgrounds. These searches are typically conducted within a given final state, characterised by the multiplicity of jets, $b$-jets, and leptons. This is useful both for scientific reasons (the SM backgrounds have an entirely different composition, and thus require a dedicated measurement, in each final state), and for political reasons (organisation of physics working groups by final state is an efficient way to parallelise search efforts). We therefore continue to assume in this paper that building an analysis within a given final state is a good goal, deferring the development of techniques that creatively use information across final states to further work.
\end{enumerate}

In this paper, we focus on the second of these problems, using our unsupervised machine learning techniques to define an \emph{anomaly score}. This is an event-by-event scalar that rates how anomalous an event is in the space of variables that the anomaly score algorithm was trained on. Our underlying assumption is that the new physics must be noticeably different from the SM background in the space constructed by the reconstructed 4-vectors and multiplicities of the final state objects.
Since our approach works on 4-vector information (which, along with particle multiplicities, are amongst the most basic set of reconstructed properties in LHC events), we believe that it is more model-independent than high-level variables that target specific kinematic configurations. 
We here provide a brief overview of the techniques that are used in our study, as well as a short summary of traditional search methods.

\subsection{Traditional methods}
\label{subsec:TradMethods}

A traditional LHC search involves constructing useful physical variables that yield some separation between signal and background. These variables are then used to perform cuts that maximise the signal while minimising the background. A set of these constraints is known as a signal region. Different signals will appear in different signal regions, meaning that each signal region must be constructed specifically for a given search, though we can expect signals that are close in parameter space to be covered by similar signal regions. Some variables that are commonly used for semi-invisible particle searches are:
\begin{itemize}
    \item $E_T^{\textrm{miss}}$: Missing energy is useful in identifying models with extra invisible particles, or anomalous production of SM neutrinos. Heavy supersymmetric particles decaying to the lightest neutralinos will yield a significantly broader $E_T^{\textrm{miss}}$ distribution than the SM background.
    
    \item $H_T$: The scalar sum of the $p_T$ of objects of interest is correlated with the energy scale of the hard process, meaning that it will generally have a broader distribution for events which produce heavy BSM particles compared to the SM events.
    
    \item $m_{\textrm{eff}}$: The scalar sum of the $p_T$ of objects of interest plus the $E_T^{\textrm{miss}}$ is a similarly useful variable, yielding a broader distribution for cases where the signal events produce much heavier particles than SM events.
    
    \item $m_T^{b,\textrm{min}}$: The transverse mass calculated from the $E_T^{\textrm{miss}}$ and the $b$-tagged jet closest in $\phi$ to the $p_T^{\textrm{miss}}$ direction is commonly used to reject events in which a $W$ boson decays via a lepton and neutrino. This is helpful to reject $t\bar{t}$ background events in searches for new particles that decay to top quarks (such as the supersymmetric top quarks we consider in our benchmark models).
\end{itemize}

\subsection{Isolation Forests}
\label{subsec:IsoFor}

As first outlined in Ref.~\cite{isolation-forest}, the Isolation Forest (IF) is an unsupervised learning algorithm that assigns each point in a dataset a value based on the ease with which it is isolated from the other points in the dataset. It is attractive due to its simple concept, linear time complexity and low memory requirement.

Given a set of data $X = \{\vec{x}_1, \vec{x}_2,~\ldots~, \vec{x}_n \}$ from a multivariate distribution, where each $\vec{x}_i$ is a vector with $d$ dimensions, one first randomly chooses a feature $k \in \{1, ..., d\}$, and a ``split value'' $p$ which lies between the maximum value and minimum value of the feature $k$. Then all $\vec{x}_i$ of the dataset with $x_{ik} < p$ are  placed in a set of points called $X_l$ while if $x_{ik} \geq p$, it is placed in a set called $X_r$. This process is repeated recursively, selecting a new $k$ each time, until \emph{one} of the following stopping conditions is met:
\begin{itemize}
    \item every data point $\vec{x}_i$ is isolated in its own set,
    \item all $\vec{x}_i$ in a given set are equal, 
    \item a limit imposed on the number of splits is reached.
\end{itemize} 
The sequence of splits generated are called \emph{trees}, and the number of splits in them is called the \emph{path length} of the tree. Each split is a \emph{node} of the tree. Nodes that do not begin or end trees are \emph{internal}, and those which do are \emph{external}. By randomly selecting batches of size $m$ from the dataset and constructing a tree for the batch, we construct what is called a $forest$. The combination of many trees in this way improves stability and performance.

\begin{figure}[!ht]
\centering
\includegraphics[scale=0.3]{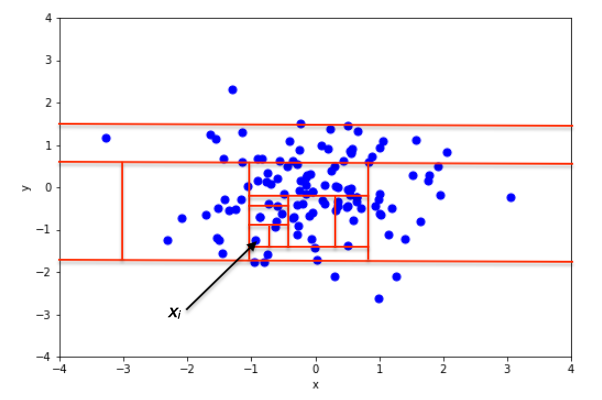}
\includegraphics[scale=0.3]{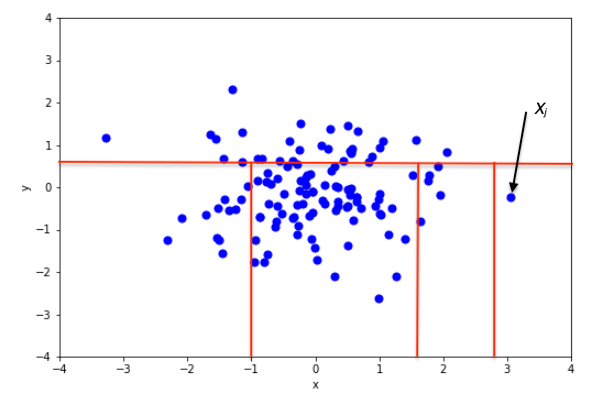}
\caption{\label{fig:isolation-forest-example} An example of two trees formed in the IF algorithm for an arbitrary 2D Gaussian distribution. Left: The isolation of a non-anomalous data point, which has path length $13$. Right: The isolation of an anomalous data point, which has path length $4$.~\cite{isolation-forest}}
\end{figure}

An anomaly is by definition an outlier, thus an anomaly should on average require a fewer number of splits to become isolated. The measure of anomalousness can then be defined via the average path length of the trees in the forest. This average path length is normalized using~\cite{isolation-forest}:
\begin{align}
	c(m) = 2H(m-1) - \left(\frac{2(m-1)}{n}\right),
\end{align}
where $n$ is the full dataset size, $m$ is the size of a randomly sampled batch, and $H(x)$ indicates the harmonic number. The anomaly score of a point $\vec{x}_i$ is then defined as
\begin{align}
	s(x,n) = 2^{-\frac{E(h(x))}{c(n)}},
	\label{eq:IF_ascore}
\end{align}
where $h(x)$ is the path length and $E(h(x))$ is the mean path length of all trees constructed for $x$. It can be seen from Eq.~\eqref{eq:IF_ascore} that $s \approx 1$ implies a high level of anomalousness (since $E(h(x))$ would be small), $s \approx 0$ indicates no anomaly at all (since $E(h(x))$ would be large). If the whole sample generates $s \approx 0.5$, we find that the entire sample is likely devoid of anomaly. For our purposes, the anomaly score has been renormalised to fall between -1 and 1, with -1 indicating no anomaly at all, and 1 indicating a high level of anomalousness. An example of two trees, one for a non-anomalous point and one for an anomalous point, in two dimensions can be found in Figure~\ref{fig:isolation-forest-example}. Note that the non-anomalous point required thirteen nodes (or splits) to isolate, while the anomalous point required only four, showing that their path lengths are vastly different.

\subsection{Gaussian mixture models}
\label{subsec:GMM}
Datasets often have subsets that share a common characteristic. Mixture models are statistical models that approximate the statistical distributions of the characteristics of such datasets. This methodology allows one to approximate a set of statistical distributions that a set of data was most likely sampled from. Specifically, Gaussian mixture models (GMMs) are an implementation of this methodology where the individual statistical distributions being fitted are Gaussian distributions~\cite{mclachlan2004finite}. The statistical distribution of the entire dataset would then build up out of these Gaussian distributions.

Let us define a set of data points as $\boldsymbol{X} = \{\vec{x}_1,~\ldots~, \vec{x}_n,~\ldots~, \vec{x}_N\}$, where each $\vec{x}_n$ is a vector with $d$ features. Let $\vec{\mu}_k$, $\vec{\Sigma}_k$ with $k = 1,\ldots,K$ be the mean vectors and covariance matrices of a chosen number ($K$) of $d$-dimensional Gaussian distributions, initialized arbitrarily. For each data point we introduce a vector of latent variables, $\vec{z}_n$ representing that it belongs to a particular Gaussian: if the $n^{\text{th}}$ data point belongs to the $k^{\text{th}}$ Gaussian we set $z_{nk} = 1$, otherwise it is zero. 

We can write the probability of observing a given data point $\vec{x}_n$ from its Gaussian as
\begin{align}
	p(\vec{x}_n|\vec{z}_n) = \prod^K_{k=1} \mathcal{N} (\vec{x}_n|\vec{\mu}_k, \vec{\Sigma}_k)^{z_{nk}},
\end{align}
where $\mathcal{N}$ denotes a Gaussian distribution. Note that this product occurs over all Gaussians but the way we have constructed the latent vector $\vec{z}_n$ suppresses all but the Gaussian $\vec{x}_n$ belongs to. Now by Bayes rule and marginalization over all $\vec{z}$ we get
\begin{align}\label{eq:logprob}
	p(\vec{x}_n) = \sum^K_{k=1} p(\vec{x}_n|\vec{z})p(\vec{z}) = \sum^K_{k=1} \pi_{k}\mathcal{N}(\vec{x}_n|\vec{\mu}_k,\vec{\Sigma}_k),
\end{align}
where we have defined a mixing parameter $\pi_k \equiv p(z_k = 1)$. These represent the probability that an arbitrary point belongs to the $k$-th mixture component (the $k$-th Gaussian), and hence the sum of $\pi_k$ over all $k$ is $1$.

We aim to maximize the probability that the observed data was sampled from the set of $K$ Gaussians ($p(\boldsymbol{X})$) by updating the parameters of those Gaussians. The log-likelihood of this probability is given by
\begin{align}
    \label{eqn:loglikelihood}
	\log\left(p(\boldsymbol{X})\right) = \sum^N_{n=1} \log\left(p(\vec{x}_n)\right) = \sum^N_{n=1} \log\left[\sum^K_{k=1}\pi_k\mathcal{N}(\vec{x}_n|\vec{\mu}_k,\vec{\Sigma}_k)\right].
\end{align}
The optimization of this function can be performed using the Expectation-Maximization (EM) algorithm. There are two steps to the EM algorithm: the expectation step (E-step) and maximization step (M-step). 

The E-step is performed by calculating the probability that each point was sampled from a particular Gaussian. This can be expressed in terms of the latent variables as $p(\vec{z}_k=1|\vec{x}_n)$, which is often referred to as the \emph{responsibility} of the distribution $k$ for a given data point $\vec{x}_n$. Using Bayes law we can write~\cite{10.2307/2984875}
\begin{align}
	p(\vec{z}_k=1|\vec{x}_n) = \frac{p(\vec{x}_n|\vec{z}_k=1)p(\vec{z}_k=1)}{\Sigma^K_{j=1}p(\vec{x}_n|\vec{z}_j=1)p(\vec{z}_j=1)} = \frac{\pi_k\mathcal{N}(\vec{x}_n|\vec{\mu}_k,\vec{\Sigma}_k)}{\Sigma^K_{j=1}\pi_j\mathcal{N}(\vec{x}_n|\vec{\mu}_j,\vec{\Sigma}_j)} \equiv \gamma(z_{nk}).
\end{align}
Once we have calculated $\gamma(z_{nk})$ for all $n$ and $k$ we can undertake the M-step to estimate the updated parameters of each Gaussian. First one calculates the number of points $N_k$ for which Gaussian $k$ is responsible
\begin{align}
	N_k = \sum^N_{n=1}\gamma(z_{nk}).
\end{align}
With this value, we update the mean of Gaussian $k$ by calculating the mean of the data points that belong to it, weighted by the responsibilities $\gamma(z_{nk})$
\begin{align}
	\vec{\mu}_k^\prime = \frac{1}{N_k}\sum^N_{n=1}\gamma(z_{nk})\vec{x}_n.
\end{align}
Similarly, the updated covariances for Gaussian $k$ are given by the covariance of the points that belong to Gaussian $k$ with the updated mean $\vec{\mu}'_k$, weighted by the responsibilities
\begin{align}
	\vec{\Sigma}_k^\prime = \frac{1}{N_k}\sum^N_{n=1}\gamma(z_{nk})(\vec{x}_n-\vec{\mu}_k^\prime)(\vec{x}_n-\vec{\mu}^\prime_k)^T.
\end{align}
Finally, the mixing parameter $\pi_k$ of Gaussian $k$ is updated by calculating the percentage of the total dataset that belong to it
\begin{align}
	\pi^\prime_k = \frac{N_k}{N}.
\end{align}
The new log-likelihood may be computed directly using Eq.~\eqref{eqn:loglikelihood} with the new parameters for each Gaussian. The process is repeated iteratively, until we see convergence of the log-likelihood (parameterized by a tolerance), or when the maximum number of epochs is reached. The anomaly score of a given data point (event) is then given by the log-probability $\log p(\vec{x}_n)$, with $p(\vec{x}_n)$ defined in Eq.~\eqref{eq:logprob}.

\subsection{Autoencoders}
\label{subsubsec:autoencoder}
Autoencoders~\cite{10.1145/1390156.1390294_autoencoder} (AEs) are a special class of neural networks where the input and output of the network are equal. This means that AEs can be trained without labels in unsupervised applications. The loss function typically is chosen to be the reconstruction loss, which is the difference between the output and input, quantified by, for example, the mean squared error on every dimension of the data. Generally, the number of hidden neurons in the neural network first decreases and then increases again, so the data needs to be squeezed in a lower dimensional representation. The lowest dimensional representation, usually in the middle of the network, is called the latent space. If the latent space dimensionality is too high, the neural network can simply learn the identity function to make the output equal to the input. When it is too low, too much information needs to be removed in order to have a good reconstruction ability. The part of the network that transforms the input to latent space representation is called the encoder, while the part of the network that transforms the latent space representation to output is called the decoder.

If the latent space dimensionality is chosen sensibly, the input data is transformed into a lower dimensional representation, which contains only the relevant information that is required for reconstruction of the original input. If an AE is trained on a dataset without any anomalies and applied on a dataset with both normal and anomalous data, the AE will have a low reconstruction loss for the normal events and a high reconstruction loss for the anomalous events. This is because the anomalous events are different from the normal events, and thus are placed in unexpected locations in the latent space by the encoder, which the decoder cannot reconstruct well. These anomalous events are then reconstructed badly. An AE can thus be used as a anomaly detector where the anomaly score of a given event is defined as the reconstruction loss of that event~\cite{10.1145/2689746.2689747_ae_anomaly}.

\begin{figure}[t]
\centering
 \includegraphics[width=0.6\textwidth]{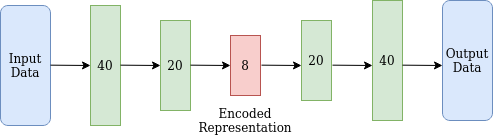}
 \includegraphics[width=0.6\textwidth]{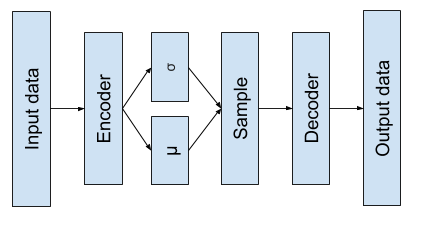}
 \caption{Schematic of our AE (top) and a VAE (bottom).}
 \label{fig:vae_and_ae}
\end{figure}

In this work, we define an AE with 5 hidden layers that have 40, 20, 8, 20, and 40 nodes, as shown in Figure~\ref{fig:vae_and_ae} (top). This shape is modelled after~\cite{hajer2018novelty}. Each layer uses a sigmoid activation function. The loss function used is a Sliced Wasserstein Distance Metric~\cite{kolouri2018sliced}. The Wasserstein Distance (sometimes referred to as ``Earth Movers Distance'') between two distributions $u$ and $v$ can be thought of as the minimum amount of energy required to transform $u$ into $v$, where the energy is defined by a cost function given by the distribution weight multiplied by the distance to move the distribution. It is a useful tool as it metrizes the energy flow between two events. The Sliced Wasserstein Distance is the Wasserstein Distance between a projection of the data onto a 1-D distribution. It has similar properties to the Wasserstein Distance metric, and is more computationally efficient. 

In addition to using the reconstruction loss as an outlier detection variable, one can also explore the latent space of an AE. If the latent space has ordering (similar events are clustered closely together in latent space) and the AE is trained to correctly reconstruct the standard model background, the latent space variables offer another representation of the standard model events. While the input space can have discontinuous and categorical data, the latent space only contains continuous data. This makes working with the latent space representation much more easy than working with 4-vector information, and one can define other outlier detection techniques on top of the latent space representation of the data. 

\subsection{Variational autoencoders}
\label{subsubsec:vae}
An AE does not have ordering in the latent space, because there is no term in the loss function that constrains the latent space. The variational autoencoder~\cite{kingma2013autoencoding_vae} (VAE), however, has this property, obtained by modifying the middle part of the neural network. The encoder outputs two numbers per latent space dimension, which represent the mean and standard deviation of a Gaussian distribution (see Figure~\ref{fig:vae_and_ae}). The decoder takes a random sample of this distribution and decodes the sample into the original input. The loss function is augmented such that the KL-divergence~\cite{Kullback59} of these Gaussians and a standard normal distribution should be as low as possible. The loss function of a VAE then consists a function that encodes the ability to reconstruct the original data point, and a KL-divergence term. The former optimises for optimal reconstruction, while the KL-divergence term forces ordering in the latent space: all input should be encoded as close to $\vec{0}$ in the latent space as possible.

During training, a balance between the two contributions to the loss function should be found: if the KL-divergence term is zero, all input is encoded to $\mathcal{N}(0, 1)$, which means there is no ability to reconstruct different points any more. The relative importance between the terms can be tuned. This was first done in Ref.~\cite{Higgins2017betaVAELB_betavae}, where it was shown that if the KL-divergence term is more important than the reconstruction loss term, one will achieve disentanglement: every latent space dimension describes a different feature in the dataset. Then, Ref.~\cite{Otten:2019hhl_bvae} showed that if the reconstruction loss term is more important than the KL-divergence term, ordering in the latent space still occurs while you can achieve a very good reconstruction loss. The relative importance of these two contributions is parameterised by $\beta$ for the reconstruction loss and $(1-\beta)$ for the KL-divergence term, and we set $\beta=3\cdot10^{-3}$ in our work. 

Using the event structure defined in Eq.~\eqref{eq:event_structure}, the reconstruction loss is chosen to consist of three components: a-mean-squared error on the number of objects $x_n$, a-mean-squared error on the regression variables ($\vec{x}_{r,i} = p_T, \eta\ \textrm{or}\ \phi$, see Eq.~\eqref{eq:event_structure}), and a categorical cross-entropy (see e.g.~\cite{murphy2013machine}) on the categorical variables $x_{c,i}$ that represent the types of the different objects in an event (see Eq.~\ref{eq:event_structure}). The total loss function of the VAE is then defined as
{\allowdisplaybreaks\begin{eqnarray}
    \mathcal{L} &= &100\beta\left(x_n - \hat{x}_n\right)^2  
\label{eq:vaeloss} \\
 && +\,  \frac{\beta}{d_r}\sum_i^{d_r}\left(x_{r,i} - \hat{x}_{r,i}\right)^2 \nonumber \\
 &&-\, \frac{10\beta}{d_c}\sum_i^{d_c}\left(x_{c,i}{\rm log}(\hat{x}_{c,i})+(1-x_{c,i}){\rm log}\left(1-\hat{x}_{c,i}\right)\right) \nonumber \\ && +\,  \left(1-\beta\right)\sum_i^{d_z}\rm{KL}\left(\mathcal{N}(\hat{\mu}_i, \hat{\sigma}_i), \mathcal{N}(0, 1)\right).\nonumber 
\end{eqnarray}}
Here, $\hat{x}_n$ the predicted number of objects, $\hat{x}_{r,i}$ the $i$-th predicted regression label, $\hat{x}_{c,i}$ the $i$-th predicted categorical label, $d_r$ the number of regression variables, and $d_c$ the dimensionality of the categorical data. The relative importance of these contributions to the loss function is indicated by $\beta$. The first three components together form the total reconstruction loss, and the last component is the KL-divergence loss term. Because the the three components for the reconstruction loss are not equally important, they are weighted with a numerical factor for the first and third line.

Regarding the VAE architecture, we have used for the encoder and decoder 3 fully-connected hidden layers, each containing 512, 256 and 128 nodes respectively for the former, and 128, 256 and 512 nodes respectively for the latter. The activation function used between the hidden nodes is the exponential linear unit (ELU)~\cite{clevert2016fast}. The latent space dimensionality is chosen to be 13. 

\begin{table}[b]
    \centering
    \begin{tabular}{c|c}
         Algorithm & Anomaly-score definition \\ \hline
         Isolation forest (IF, Section~\ref{subsec:IsoFor}) & Mean path length (Eq.~\eqref{eq:IF_ascore}) \\
         \hline
         Gaussian mixture model (GMM, Section~\ref{subsec:GMM}) & Log probability (log of Eq.~\eqref{eq:logprob}) \\
         \hline
         Static autoencoder (AE, Section~\ref{subsubsec:autoencoder}) & Sliced Wasserstein Distance~\cite{kolouri2018sliced} \\
         \hline
         Variational autoencoder (VAE, Section~\ref{subsubsec:vae}) & \begin{tabular}{@{}c@{}}Reconstruction loss\\
         (first three lines of Eq.~\eqref{eq:vaeloss}) \end{tabular} \\
    \end{tabular}
    \caption{Summary of the considered ML algorithms and the definition of their anomaly scores. }
    \label{tab:ascore_defn}
\end{table}
\subsection{Concluding remarks and combinations of anomaly detection methods}
\label{subsec:combinations}

Table \ref{tab:ascore_defn} summarizes the anomaly scores for each algorithm that we have discussed. Since the VAE transforms the events into a lower-dimensional continuous space, it is believed that the other outlier detection techniques can find outliers more easily in the latent space of the VAE. Therefore, besides for exploring the above-discussed algorithms individually, we will also apply the IF, GMM, and the AE on the latent space of the VAE.

Moreover, if the anomaly scores yielded from each algorithm are minimally correlated with each other, there is information to be gained from combining them. In this paper we explore AND, OR, product, and averaging combinations. For a given event, let the anomaly score normalised to uniform background efficiency be $x_i$ where $i$ denotes the anomaly score algorithm. The AND anomaly score combination is given by $x^{\text{AND}} = \textrm{min}(x_i)$ for a given event, whereas OR anomaly score combination is given by $x^{\text{OR}} = \textrm{max}(x_i)$. The product combination is given by $x^{\text{product}} = \prod_i x_i$, and the average combination is given by $x^{\text{average}} = \frac{1}{N} \sum_i x_i$, where $N$ is the number of algorithms being used, in this case, 4 (IF, GMM, AE, and the VAE). 

\section{Results}
\label{sec:results}

In the following, we investigate two different ways of applying our techniques to particle physics data. In both cases, we train the techniques after applying a minimal pre-selection of:

\begin{itemize}
    \item $E_T^{\textrm{miss}} \geq 150 \text{GeV}$,
    \item $\geq 4$ jets,
    \item $H_T \geq 600 \text{GeV}$.
\end{itemize}

In our first approach, we train the IF, GMM, AE and VAE directly on the 4-vector representation of the events that pass the pre-selection, and compare the relative performance of each technique. We then also assess the performance of the various combination techniques described in Section~\ref{subsec:combinations}. The full list of input variables is $(E_T, \phi)_{\textrm{miss}}, (E, p_T, \eta,  \phi)_{\textrm{jets}}$, $(E, p_T, \eta, \phi)_{\textrm{bjets}}, (E, p_T, \eta, \phi)_{\textrm{leptons}}, (E, p_T, \eta, \phi)_{\textrm{photons}}$. Here, leptons can be positively or negatively charged electrons or muons. In our second approach, we train the VAE in the same way, but apply the IF, GMM and AE algorithms to the latent space variables defined by the VAE. The combination of techniques applied to these anomaly score methods end up providing the optimum results, as we will see in what follows.
\subsection{Comparison of techniques with training on original 4-vectors}

\begin{figure}[!h]
 \begin{minipage}[t]{0.5\textwidth}
 \centering  
 \includegraphics[width=\linewidth]{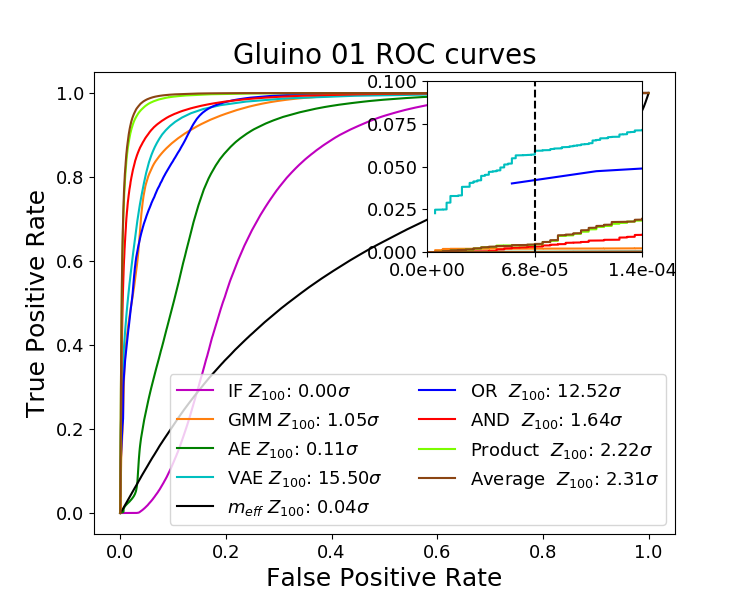}
 \includegraphics[width=\linewidth]{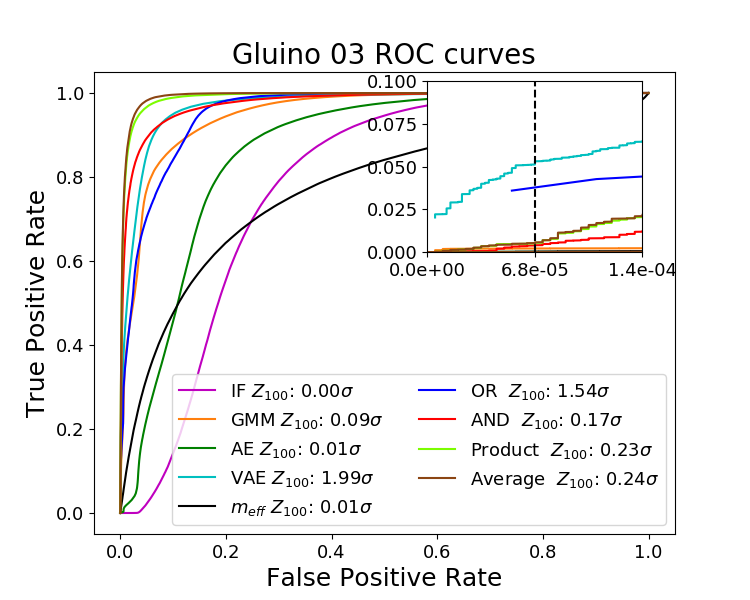} 
 \includegraphics[width=\linewidth]{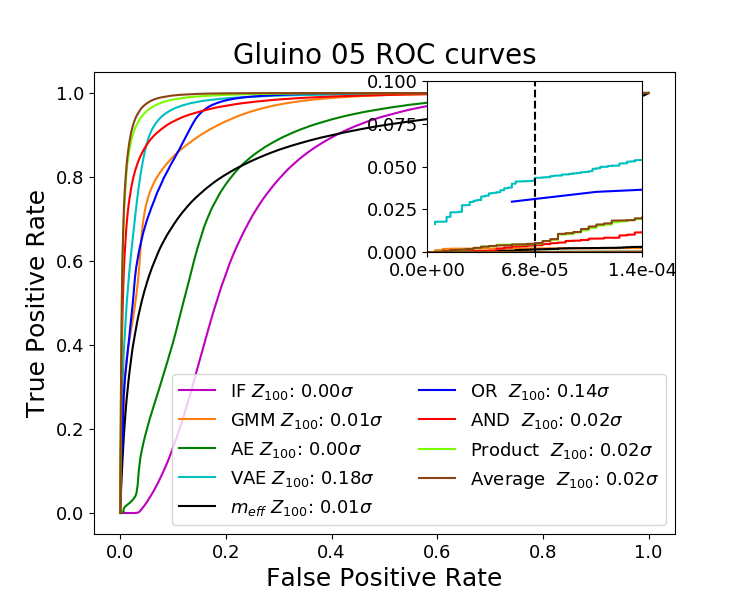} 
 \end{minipage}
\begin{minipage}[t]{0.5\textwidth}
 \centering  
 \includegraphics[width=\linewidth]{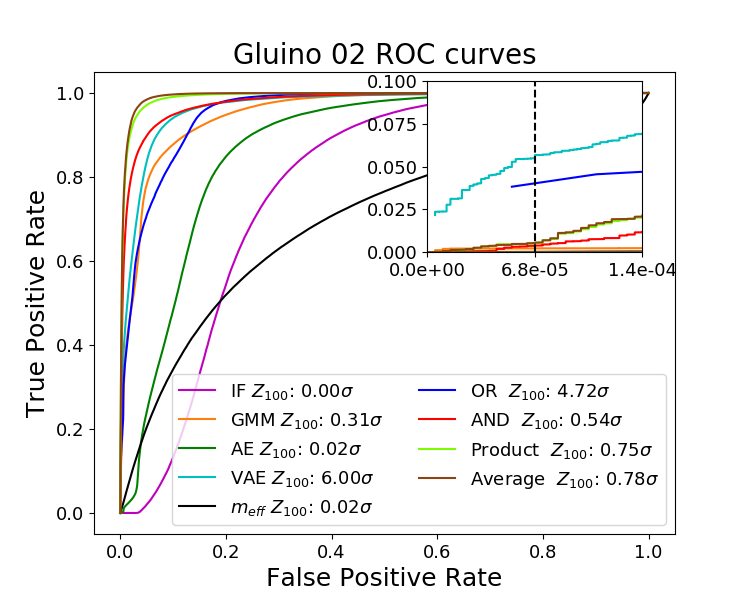}
 \includegraphics[width=\linewidth]{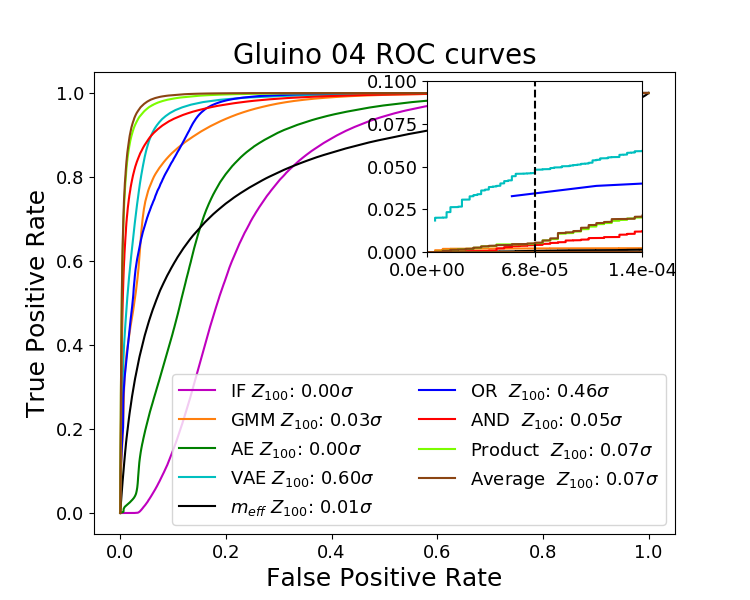} 
 \includegraphics[width=\linewidth]{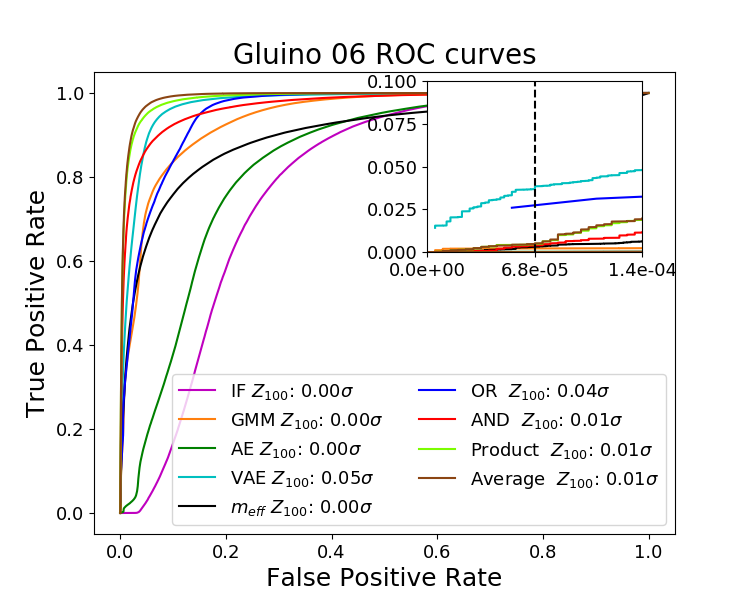} 
\end{minipage}
 \caption{ROC curves for the gluino signals (Table~\ref{tab:bsm-proc}) for the algorithms applied on 4-vector representations, with on the horizontal (vertical) axis the false-positive (true-positive) rate. The ROC curves of IF, GMM, AE and VAE (Table~\ref{tab:ascore_defn}) are shown in pink, orange, dark green and cyan respectively. The effective mass $m_{\rm eff}$ is shown in black, and combinations of the models are shown in blue (OR), red (AND), light-green (Product) and brown (Average). The inset panel shows the performance of the the curves for $B=100$, indicated by the dashed black line. }
 \label{fig:ROC_4vector_gluino}
\end{figure}
\begin{figure}[h]
\begin{minipage}[t]{0.5\textwidth}
 \centering  
 \includegraphics[width=\linewidth]{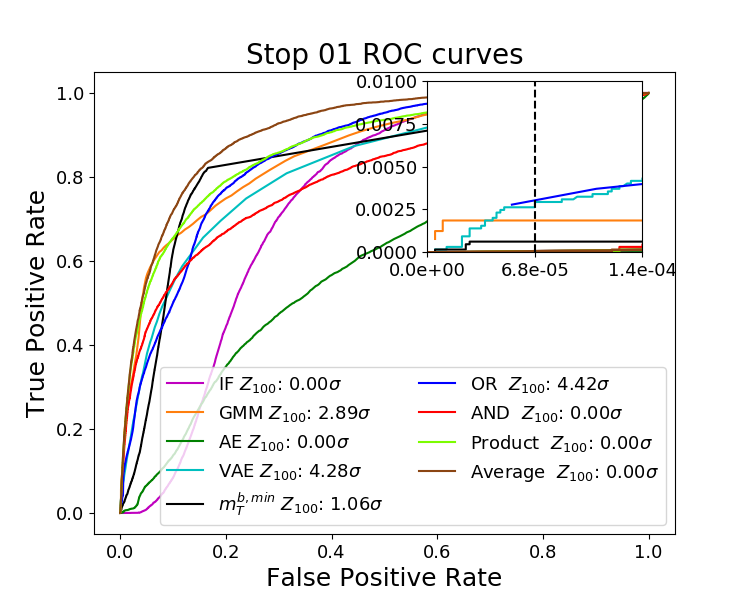}
 \includegraphics[width=\linewidth]{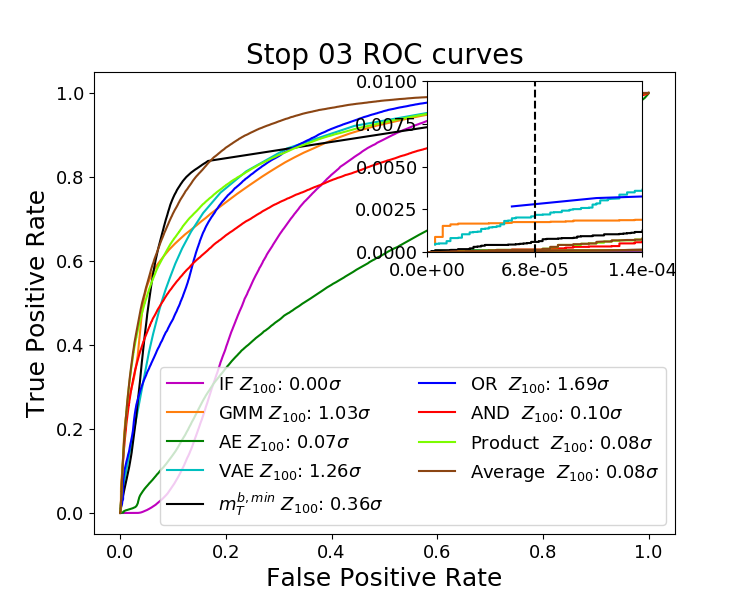} 
 \end{minipage}
\begin{minipage}[t]{0.5\textwidth}
 \centering  
 \includegraphics[width=\linewidth]{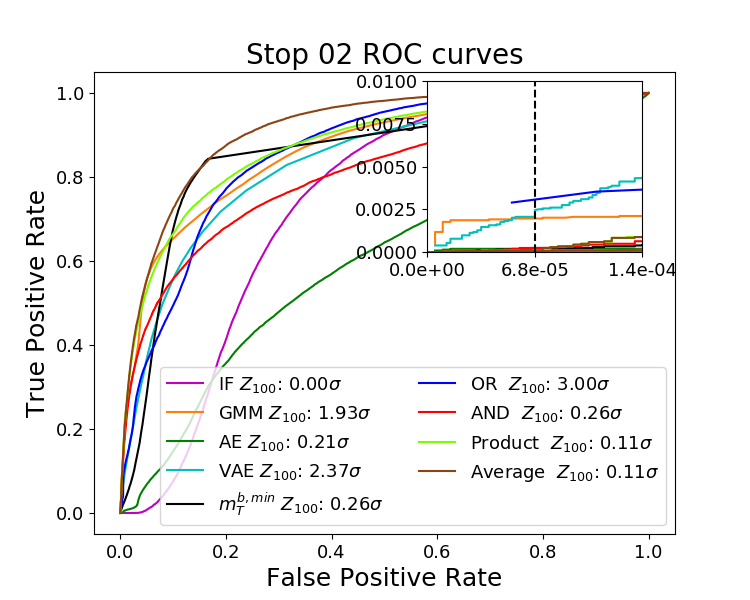}
 \includegraphics[width=\linewidth]{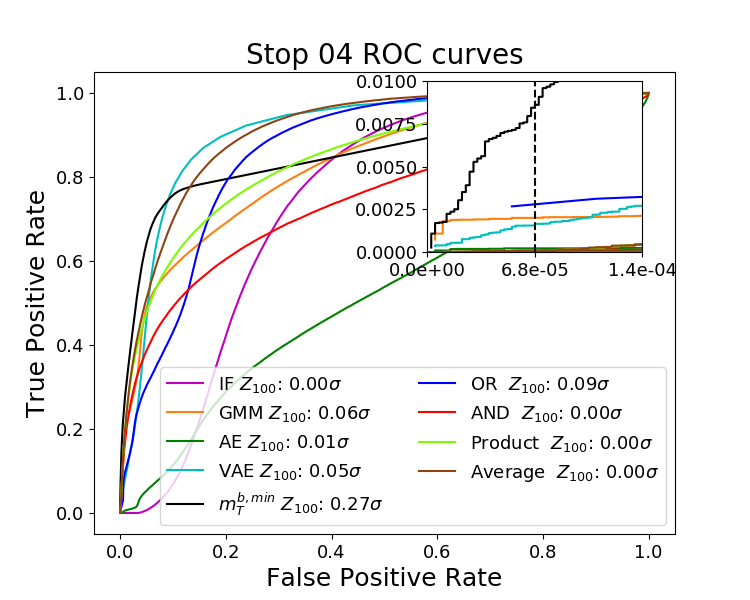} 
\end{minipage}
    \caption{ROC curves for the stop signals (Table~\ref{tab:bsm-proc}) for the algorithms applied on 4-vector representations. For further information see Figure~\ref{fig:ROC_4vector_gluino}. The physical variable that is used here is $m_{\rm T}^{\rm b,min}$.}
    \label{fig:ROC_4vector_stop}
\end{figure}

Let us first deal with the case of applying the techniques directly on the 4-vectors of the selected events. Having used each technique to define an anomaly score, we show in Figure~\ref{fig:ROC_4vector_gluino} the ROC curves for each algorithm that result from applying anomaly score cuts to the gluino signals. Given a particular selection on the anomaly score (see Table~\ref{tab:ascore_defn}), the true positive result is calculated as the proportion of signal events to the right of the cut, whilst the false positive rate is calculated as the proportion of background events to the right of the cut. We also quote significance values, calculated for the choice of anomaly score cut that results in 100 background events being selected in 36 fb$^{-1}$ of LHC data. The inset on each panel of the figures shows the ROC curves near this region, where the black dashed line indicates the 100 background event cut. 
Significance values are calculated using 
\begin{equation}
\label{eq:significance}
Z_{B} = \frac{S}{\sqrt{S+B+(\sigma_BB)^2}},
\end{equation}
\clearpage 
\noindent where $S$ is the number of signal events, $B$ is the number of background events (100 in this case), and $\sigma_B$ is the assumed systematic uncertainty. We use $Z_{100}$ to compare the algorithms in what follows. We start our discussion by assuming zero systematic uncertainty ($\sigma_B = 0$ (figures \ref{fig:ROC_4vector_gluino}, \ref{fig:ROC_4vector_stop}, \ref{fig:ROC_latent_gluino}, and \ref{fig:ROC_latent_stop}). Plots comparing the significance values yielded from the 4-vector and latent space representations by performing a cut at 100 background events with and without a 15\% relative systematic uncertainty are displayed and discussed in Section~\ref{sec:summary}.

In Figure~\ref{fig:ROC_4vector_gluino} we compare the performance of the IF, GMM, AE, VAE, the effective mass (defined below), and the combinations detailed in Section~\ref{subsec:combinations} on various gluino signals, the details of which are contained in Table~\ref{tab:bsm-proc}. The effective mass is defined as $m_{\textrm{eff}} = E_T^{\textrm{miss}} + \sum_{\textrm{jets}}{p_T}$ and is a common discriminating variable in conventional gluino searches. Using $Z_{100}$, we see that the VAE provides the strongest separation between signal and background for all gluino signal models. The OR combination also gives a good separation, though it's not as effective as the VAE. The relatively poor performance of the IF algorithm can be explained by the fact that dividing the space by placing cuts in the space of the 4-vector components does not obviously isolate outlying events, since the anomalies are more likely to appear in non-trivial functions of the four-vector components. The AE and VAE perform better because their attempts to reproduce the structure of the background events involve defining non-linear functions of the input variables that do not then generalise well to the case of the signal events. Past the $100$-background-event cut, the product and average combinations excel based off the area under the curve for each. Their relatively poor performance at the 100 event cut is due to the poor performance of the IF, GMM, and AE in these low background efficiency regions. For models with a higher gluino mass, our techniques lose discovery and exclusion potential despite the more anomalous kinematics of these models. This is caused by the reduced production cross-section in each case, resulting in a smaller value for $S$. It is possible that our anomaly score could be supplemented by traditional kinematic selections, and we return to this point in the next subsection.

Figure~\ref{fig:ROC_4vector_stop} displays the ROC curves that result from applying each algorithm applied to the various stop signals, the details of which are contained in Table~\ref{tab:bsm-proc}. The algorithms are compared to  $m_T^{b, \textrm{min}} = \sqrt{2p_T^bE_T^{\textrm{miss}}[1-\cos\Delta\phi(p_T^b,p_T^{\textrm{miss}})]}$, which is a common discriminating variable for stop signals. The first of these signals is particularly difficult to differentiate from the background, since it is kinematically very similar to the dominant background processes. Using $Z_{100}$, the OR combination (marginally) yields the strongest separation between signal and background. The VAE also provides a fair separation for all signals, while the GMM is comparable, although definitely a poorer anomaly detector for the stop case. The IF and the AE give much worse performance, which is easy to rationalise in the case of the IF using a similar argument to that provided for the gluino results. The AE is less flexible than the VAE, which explains its poor performance. It remains to be seen whether a more complex AE would improve the performance (which we will not pursue here by virtue of having defined the VAE). Surprisingly, the values for $Z_{100}$ show the most promising results for the lower stop mass case despite being the most standard model-like. The reason that this happens is because of its cross section, which is significantly higher than those of the other tested stop scenarios. For higher stop masses it is clear that our techniques would not deliver discovery or exclusion potential despite the more anomalous kinematics, which is again driven by the falling production cross-section as the stop mass increases. Ultimately, this results from the fact that the anomaly score alone is not an effective discriminant between the signal and background for stop models. For the stop 04 signal, the variable $m_T^{b, \textrm{min}}$ outperforms the algorithms, which suggests there is information contained in this variable that is not being picked up by the algorithms.

\subsection{Comparison of techniques with training on latent space variables}
\begin{figure}[ht]
 \begin{minipage}[t]{0.5\textwidth}
 \centering  
 \includegraphics[width=\linewidth]{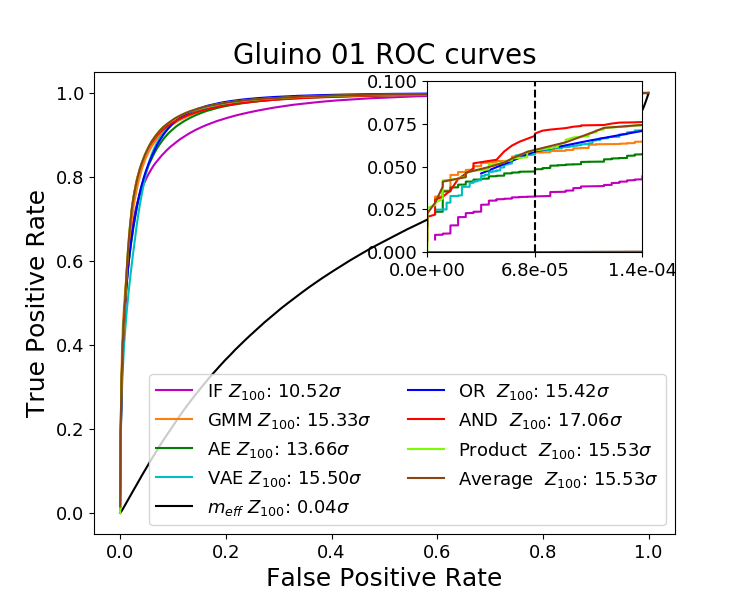} 
 \includegraphics[width=\linewidth]{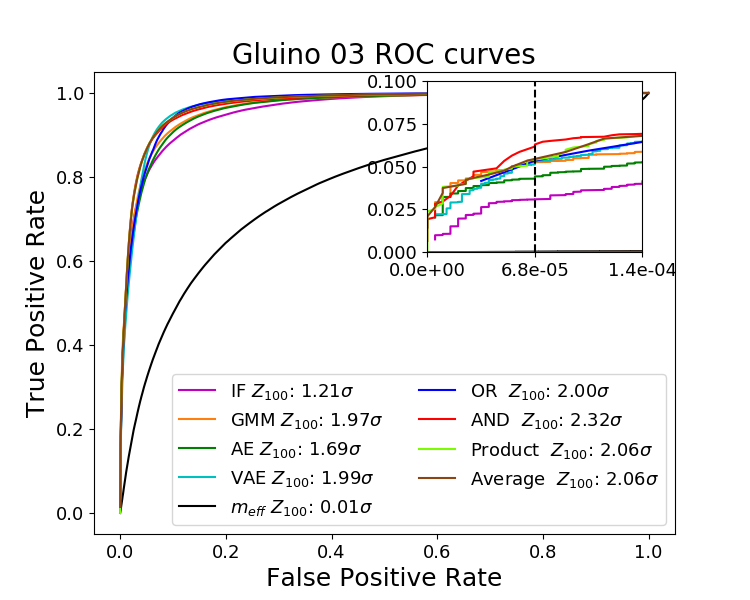} 
 \includegraphics[width=\linewidth]{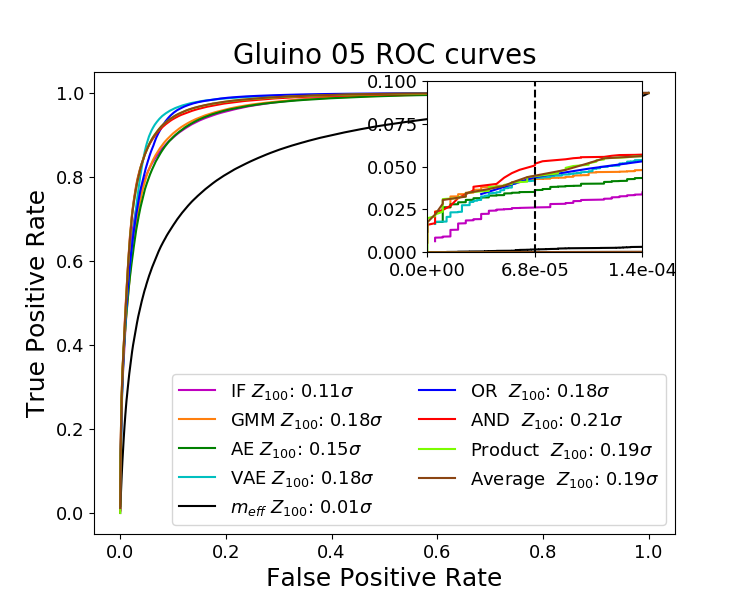} 
 \end{minipage}
\begin{minipage}[t]{0.5\textwidth}
 \centering  
 \includegraphics[width=\linewidth]{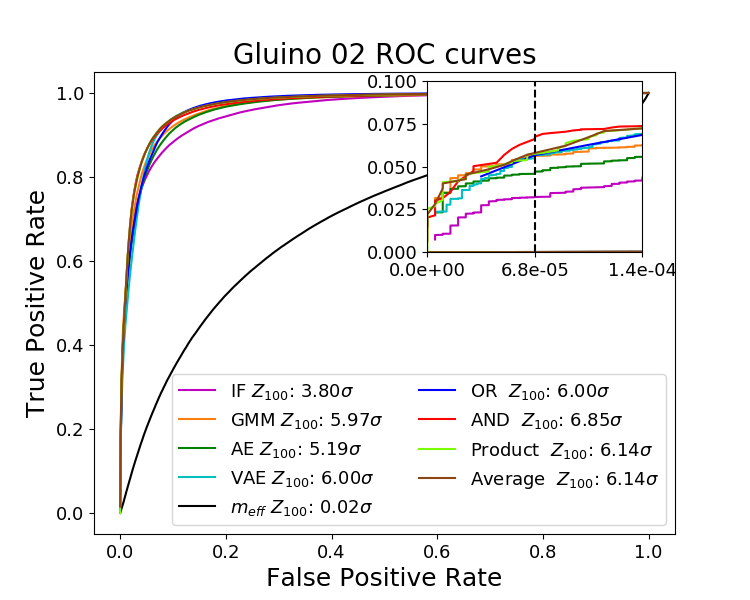} 
 \includegraphics[width=\linewidth]{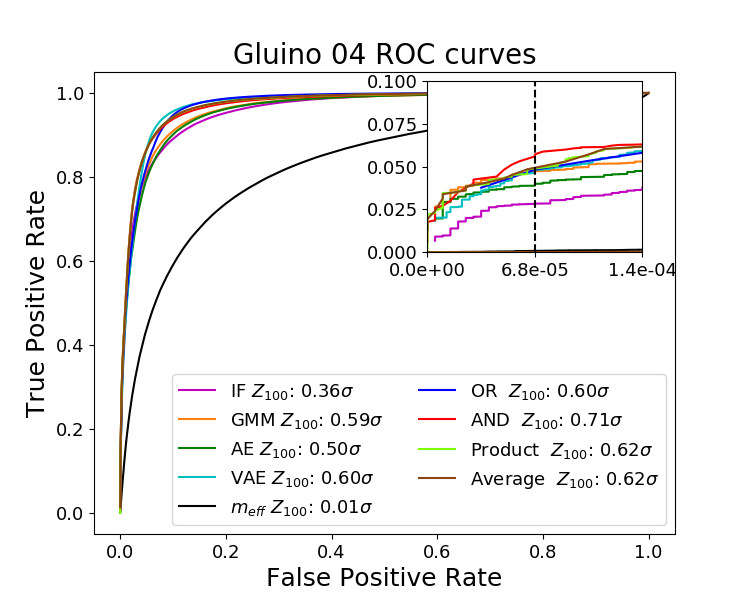} 
 \includegraphics[width=\linewidth]{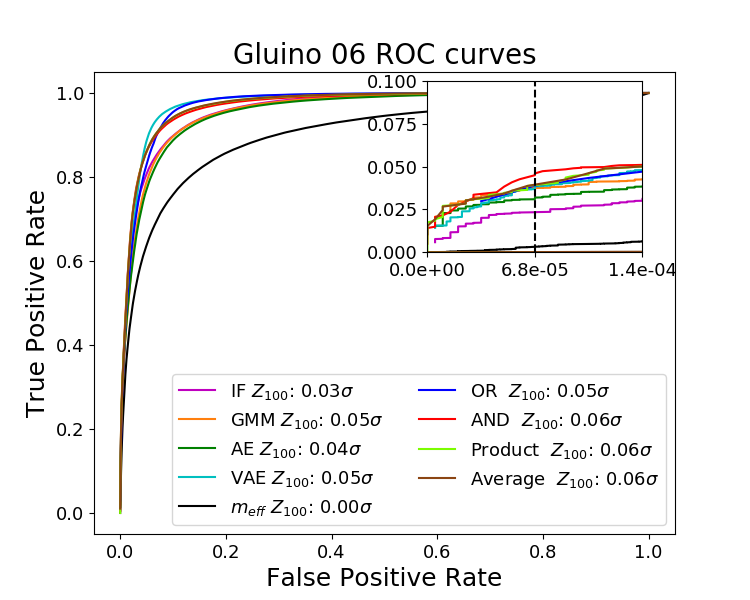} 
\end{minipage}
 \caption{ROC curves for the gluino signals (Table~\ref{tab:bsm-proc}) for the algorithms applied on latent space representations. For further information see Figure~\ref{fig:ROC_4vector_gluino}.}
 \label{fig:ROC_latent_gluino}
\end{figure}

\begin{figure}[ht]
 \begin{minipage}[t]{0.5\textwidth}
 \centering  
 \includegraphics[width=\linewidth]{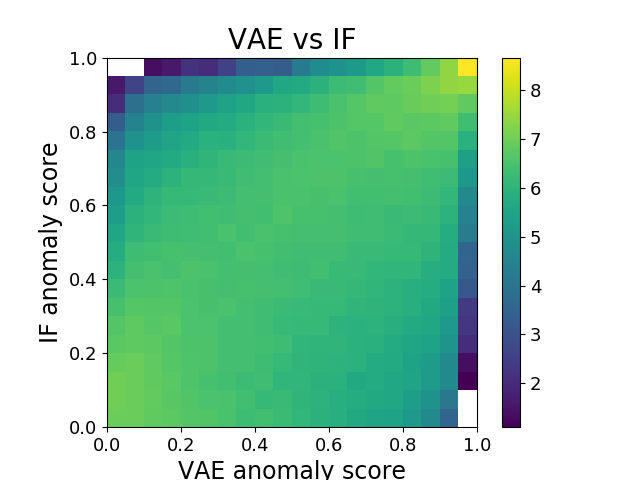}
 \includegraphics[width=\linewidth]{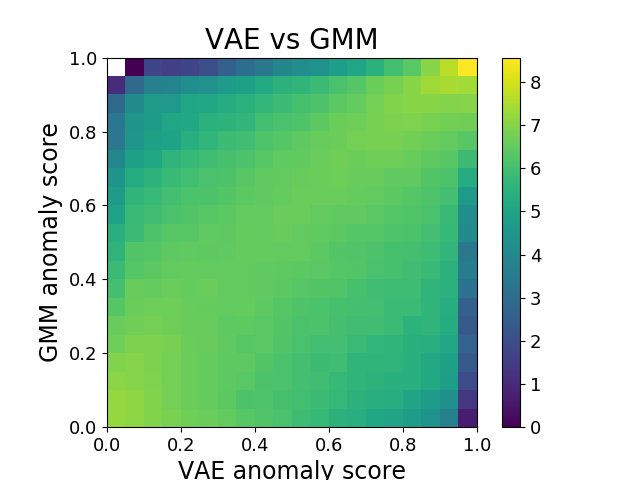}
 \includegraphics[width=\linewidth]{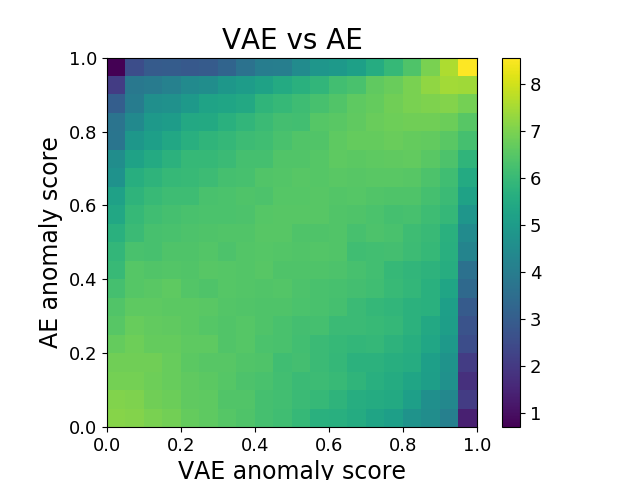} 
 \end{minipage}
\begin{minipage}[t]{0.5\textwidth}
 \centering  
 \includegraphics[width=\linewidth]{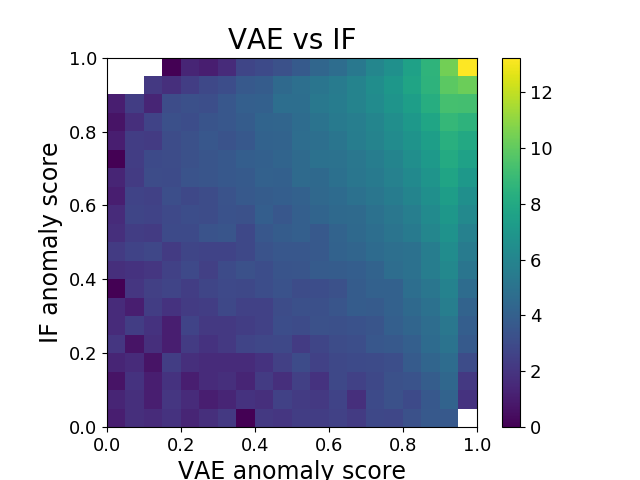}
 \includegraphics[width=\linewidth]{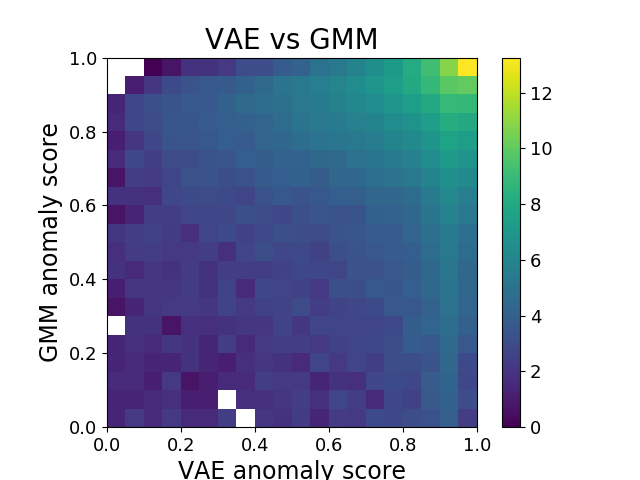}
 \includegraphics[width=\linewidth]{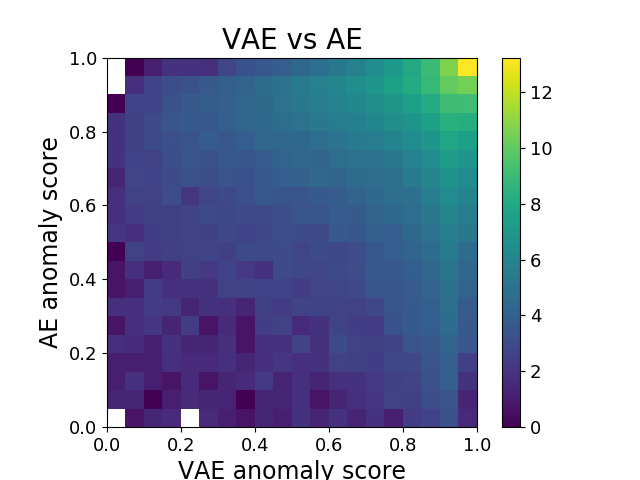} 
\end{minipage}
    \caption{2D correlation plots of various anomaly scores with normalised background efficiency compared to the VAE loss for the background (left) and gluino 01 signal (right). The colour coding represents $\log{N}$, with $N$ being the number of events.}
    \label{fig:ascore_correlations}
\end{figure}

Let us now consider the approach of training the IF, GMM and AE on the latent space representation of the SM events obtained from the VAE. The input variables for our VAE in this approach are the same as those in the previous section, and we continue to use the reconstruction loss to define the VAE anomaly score. However, the IF, GMM, and AE are trained on the 13 latent space variables defined by the VAE, which are non-linear functions of the original 4-vector variables, supplemented by the reconstruction loss of the VAE. 

In Figure~\ref{fig:ROC_latent_gluino}, we show the ROC curves for the gluino signals, which demonstrate that the performance of our non-VAE techniques has now improved dramatically in each case. The effective mass has been left in these plots for further comparison. However, we see that their performance still does not exceed that of the VAE.

When compared to the effective mass $m_{\textrm{eff}} = E_T^{\textrm{miss}} + \sum_{\textrm{jets}}{p_T}$, all anomaly score definitions outperform it by a considerable margin. Once more, we provide significance values that are calculated for an anomaly score cut that leaves 100 background events in the selected sample, assuming 36 fb$^{-1}$ of LHC data. We can see that the AND combination anomaly score outperforms all other anomaly scores, though the OR, Product and Average combinations also perform at least as well as the VAE at the 100 event cut. This can be explained by the observation that the anomaly scores yielded from the VAE and the other algorithms are minimally correlated. This may be observed in Figure~\ref{fig:ascore_correlations}, where a strong correlation would show up as a diagonal yellow line that goes from $0$ to $1$. Since this is not seen, this implies that there is further information to be gained by performing these combinations.

\begin{figure}[t]
\includegraphics[width=\linewidth]{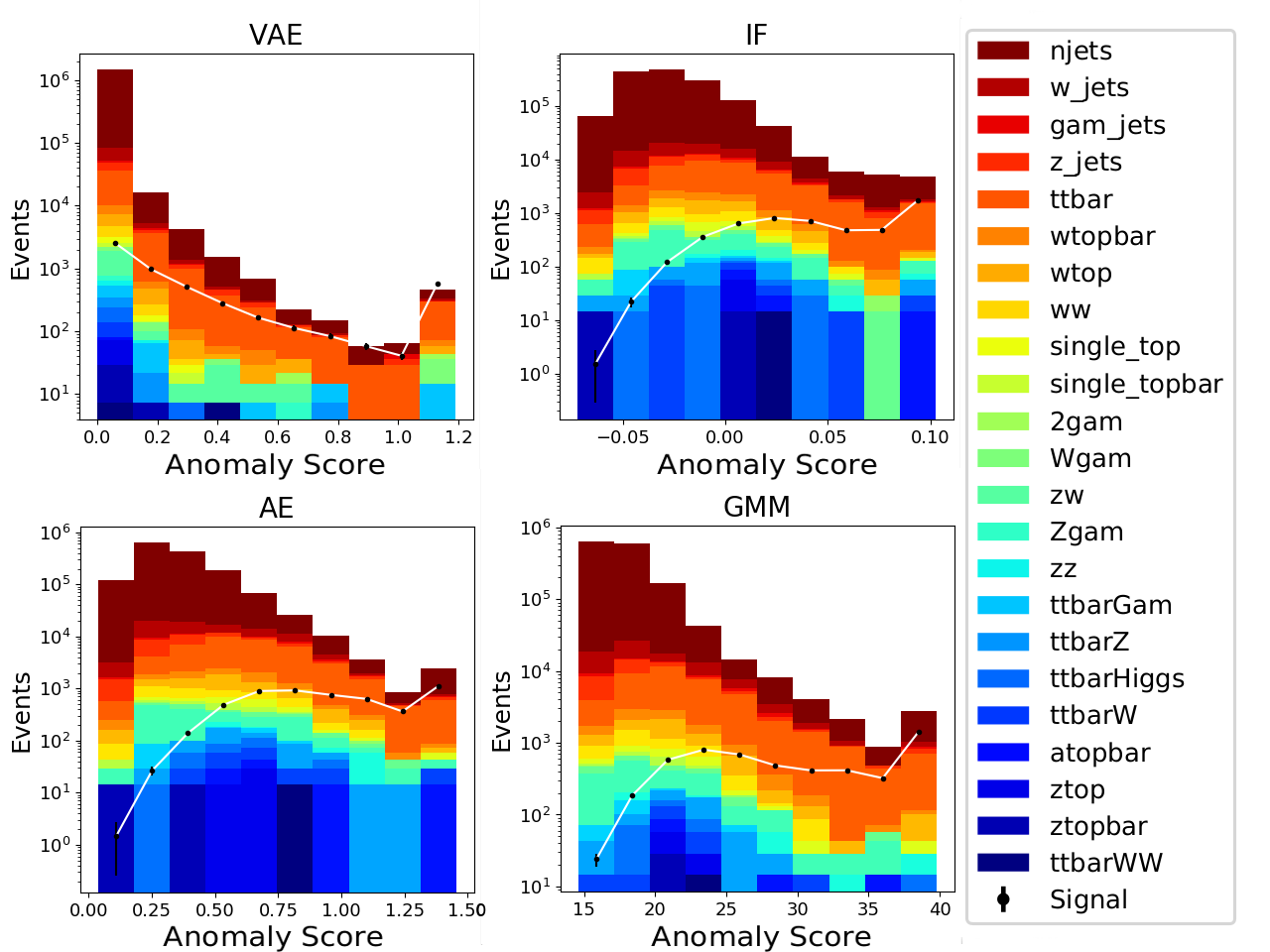} 
    \caption{Anomaly score histograms derived from various algorithms for Gluino 01. The horizontal axis shows the anomaly score, and the histogram counts the number of events normalized to 36 fb$^{-1}$ in each bin. The various colours indicate different backgrounds, while the black data points show the signal. }
    \label{fig:ascore_gluino}
\end{figure}
In Figure~\ref{fig:ascore_gluino}, we show histograms of the anomaly score itself, for both the SM background and Gluino 01. The signal is plotted separately from the background in order to better show the difference in the shape of the anomaly score distribution for each algorithm. The final bin is an overflow bin containing all events to the right of it, chosen as the first bin containing less than 10 events. These plots show a clear separation between background and signal, with the signal being more clustered in the high anomaly score region as expected. 

\begin{figure}[t]
 \begin{minipage}[t]{0.5\textwidth}
 \centering  
 \includegraphics[width=\linewidth]{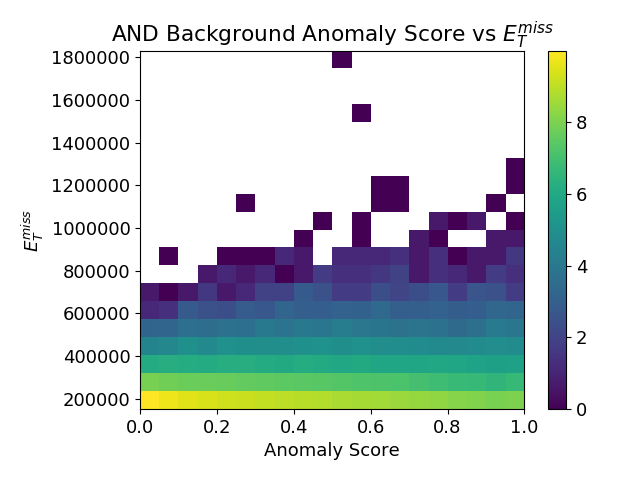}
 \includegraphics[width=\linewidth]{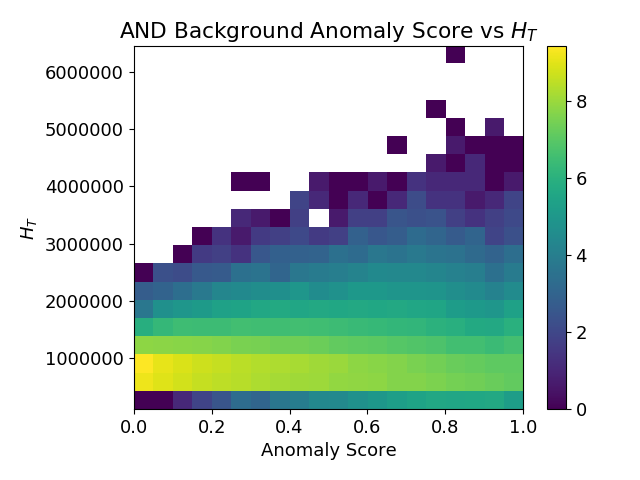}
 \includegraphics[width=\linewidth]{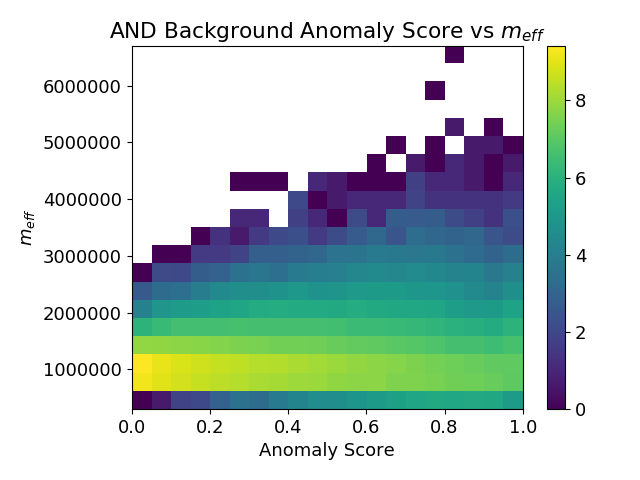}
 \end{minipage}
\begin{minipage}[t]{0.5\textwidth}
 \centering  
 \includegraphics[width=\linewidth]{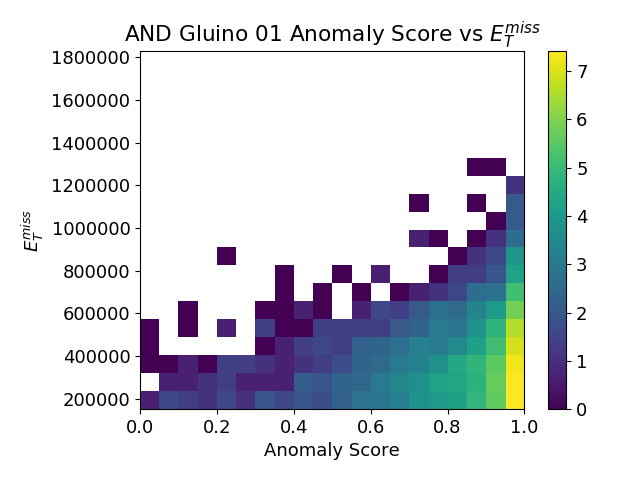}
 \includegraphics[width=\linewidth]{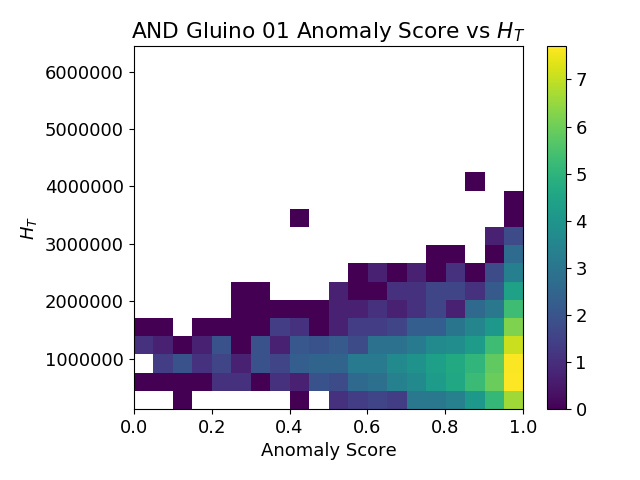}
 \includegraphics[width=\linewidth]{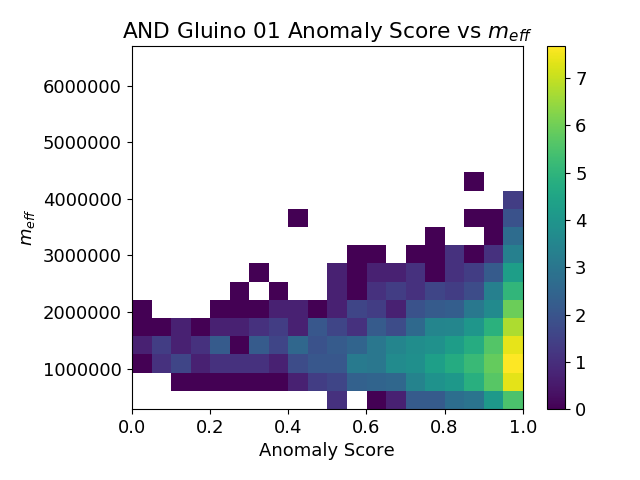}
\end{minipage}
 \caption{2D histograms associated with Gluino 01 for background (left) and signal (right). Various physical variables are plotted on the y-axis, with the anomaly score generated from the AND combination applied in the latent space on the x-axis. The z-axis is $\log{N_{\text{EVENTS}}}$}
 \label{fig:2D_IF_gluino}
\end{figure}

Figure \ref{fig:2D_IF_gluino} displays 2D correlation plots comparing the AND anomaly score to various physical variables for Gluino 01. There is minimal apparent correlation between the anomaly score and any of the physical variables that are shown in Figure~\ref{fig:2D_IF_gluino}. This implies that one could use the anomaly score as the first selection of an LHC analysis (including adding it to the high-level trigger menu), and then use conventional variables to enhance sensitivity to particular signals in the usual way. This hybrid approach reintroduces model-dependence through the later kinematic selections, but starts with very few signal assumptions. 
\begin{figure}[ht]
\begin{minipage}[t]{0.5\textwidth}
 \centering  
 \includegraphics[width=\linewidth]{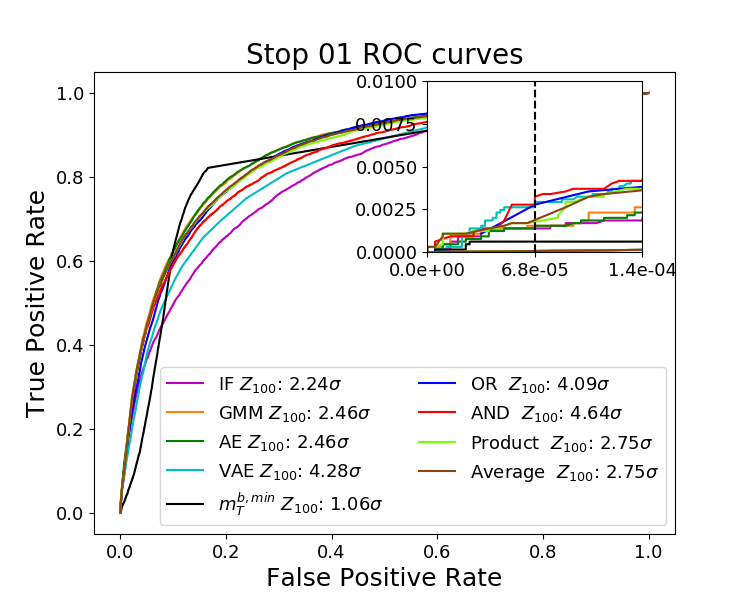}
 \includegraphics[width=\linewidth]{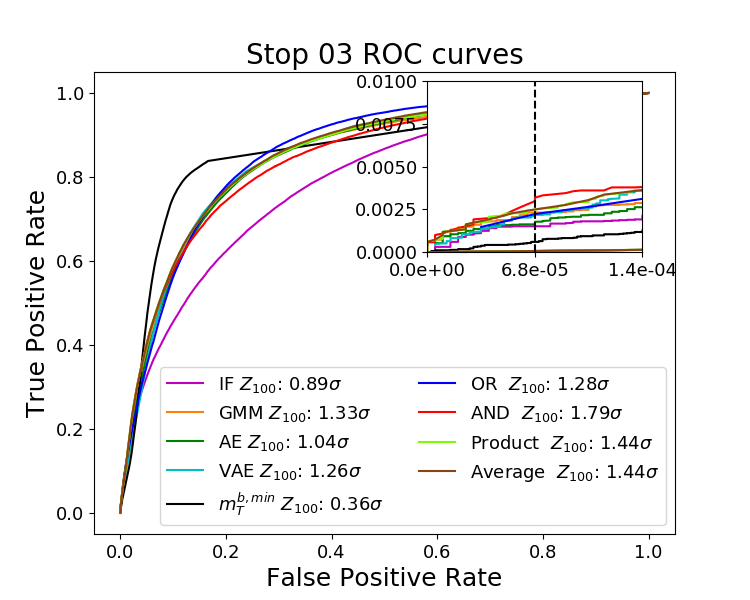}
 \end{minipage}
\begin{minipage}[t]{0.5\textwidth}
 \centering  
 \includegraphics[width=\linewidth]{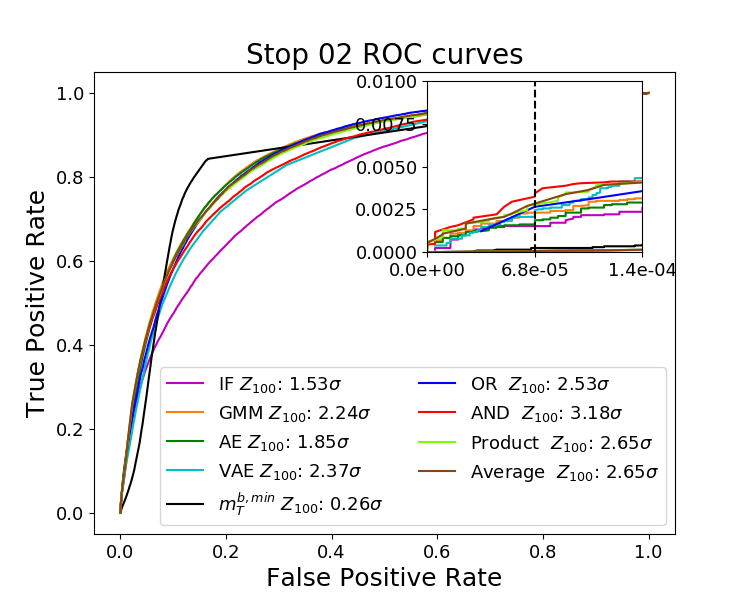}
 \includegraphics[width=\linewidth]{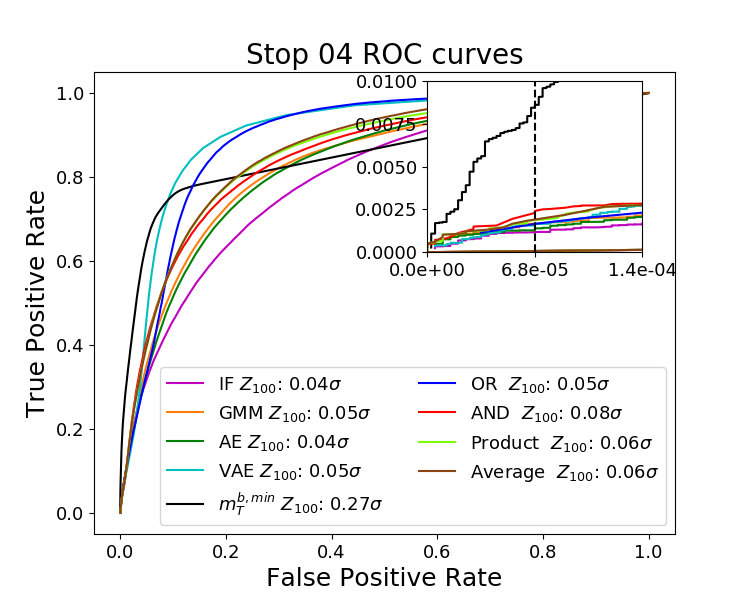}
\end{minipage}
    \caption{ROC curves for the stop signals (Table~\ref{tab:bsm-proc}) for the algorithms applied on latent space representations. For further information see Figure~\ref{fig:ROC_4vector_gluino}}
    \label{fig:ROC_latent_stop}
\end{figure}

Figure \ref{fig:ROC_latent_stop} displays ROC curves for the same algorithms being used on latent-space variables for the various stop signals. The significance numbers show promise, but it remains hard to isolate the stop signal using only a selection on the anomaly score. The IF, GMM, AE, and combination methods are much more effective when applied to the latent space. Again, the AND combination outperforms all other anomaly score definitions, except for the stop 04 signal, for which the traditional variable $m_T^{b, \textrm{min}}$ is the most effective. This suggests that, again, there is information contained in this variable that is not being picked up by the ML algorithms.

\begin{figure}[ht]
\includegraphics[width=\linewidth]{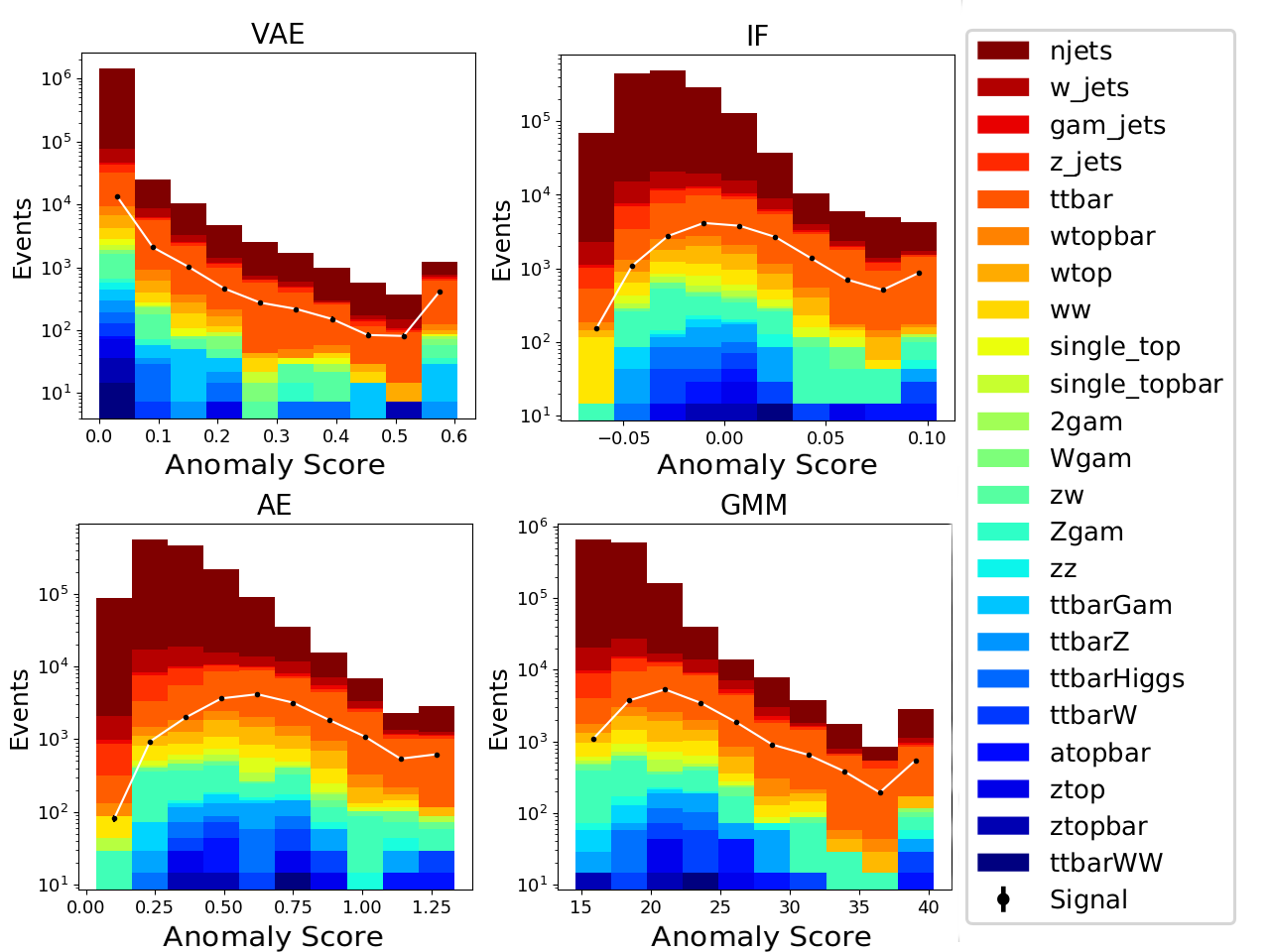}

    \caption{Anomaly score histograms derived from various algorithms for Stop 01.}
    \label{fig:ascore_stop}
\end{figure}

Figure \ref{fig:ascore_stop} displays the anomaly-score histograms for the background and Stop 01. The background and signal shapes appear much more similar than in the gluino case, though there is a tendency to push the signal to the right. Higher stop masses appear more anomalous but have a lower cross section and are thus more difficult to see at a 100 background event cut. Nevertheless, it remains the case that a loose anomaly score selection at the high-level trigger level is a well-motivated starting point for a subsequent analysis, with common kinematic variables remaining uncorrelated with the anomaly score obtained via the AND combination. 

\subsection{Summary and the inclusion of a systematic uncertainty}
\label{sec:summary}
\begin{figure}[t]
 \includegraphics[width=\linewidth]{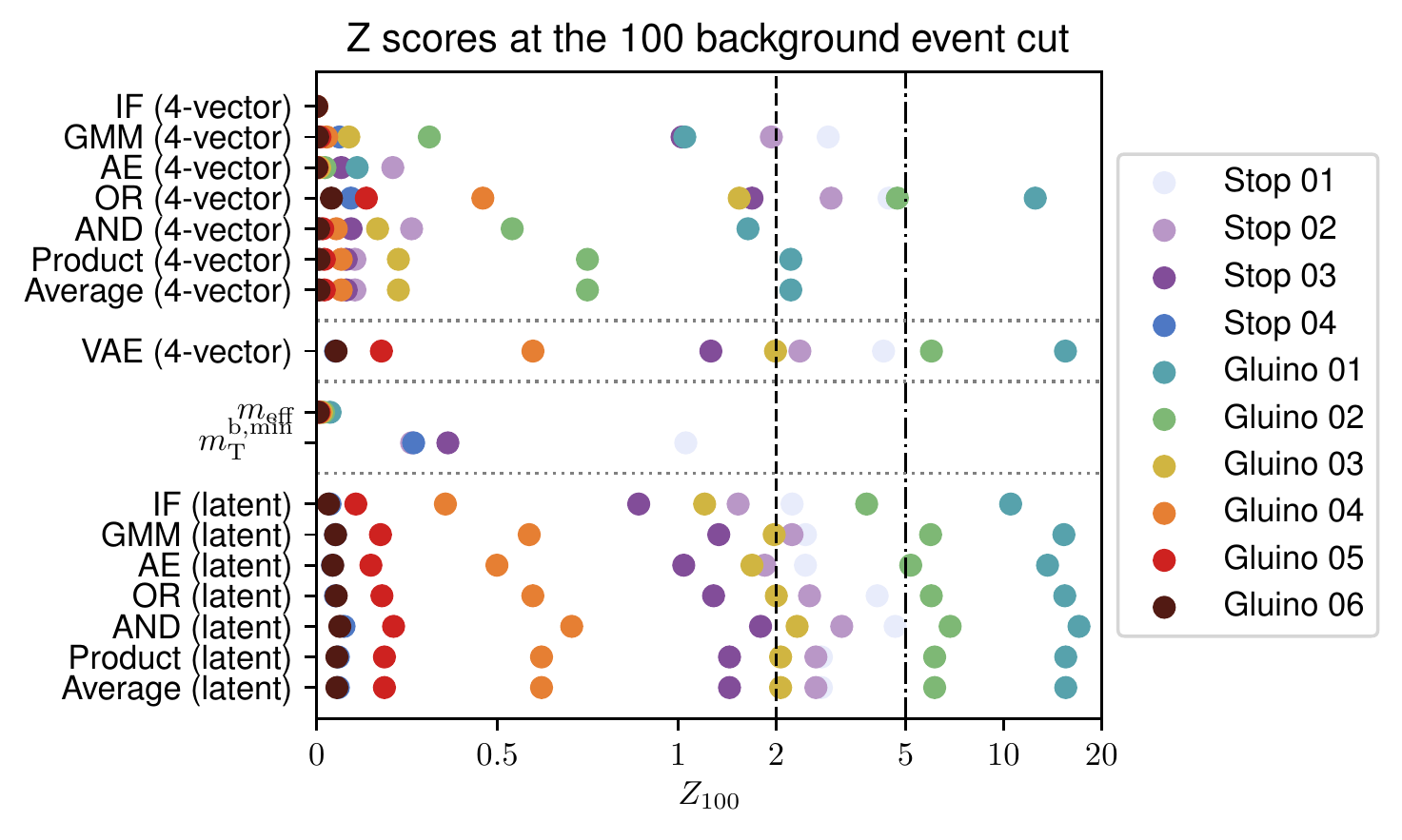}
 \caption{$Z_{100}$ yielded from various algorithms applied to 4-vector components and latent space representations. See Table~\ref{tab:bsm-proc} for the signal definitions, and Table~\ref{tab:ascore_defn} for the definitions of the algorithms. }
  \label{fig:sigma}
\end{figure}

\begin{figure}[b]
 \includegraphics[width=\linewidth]{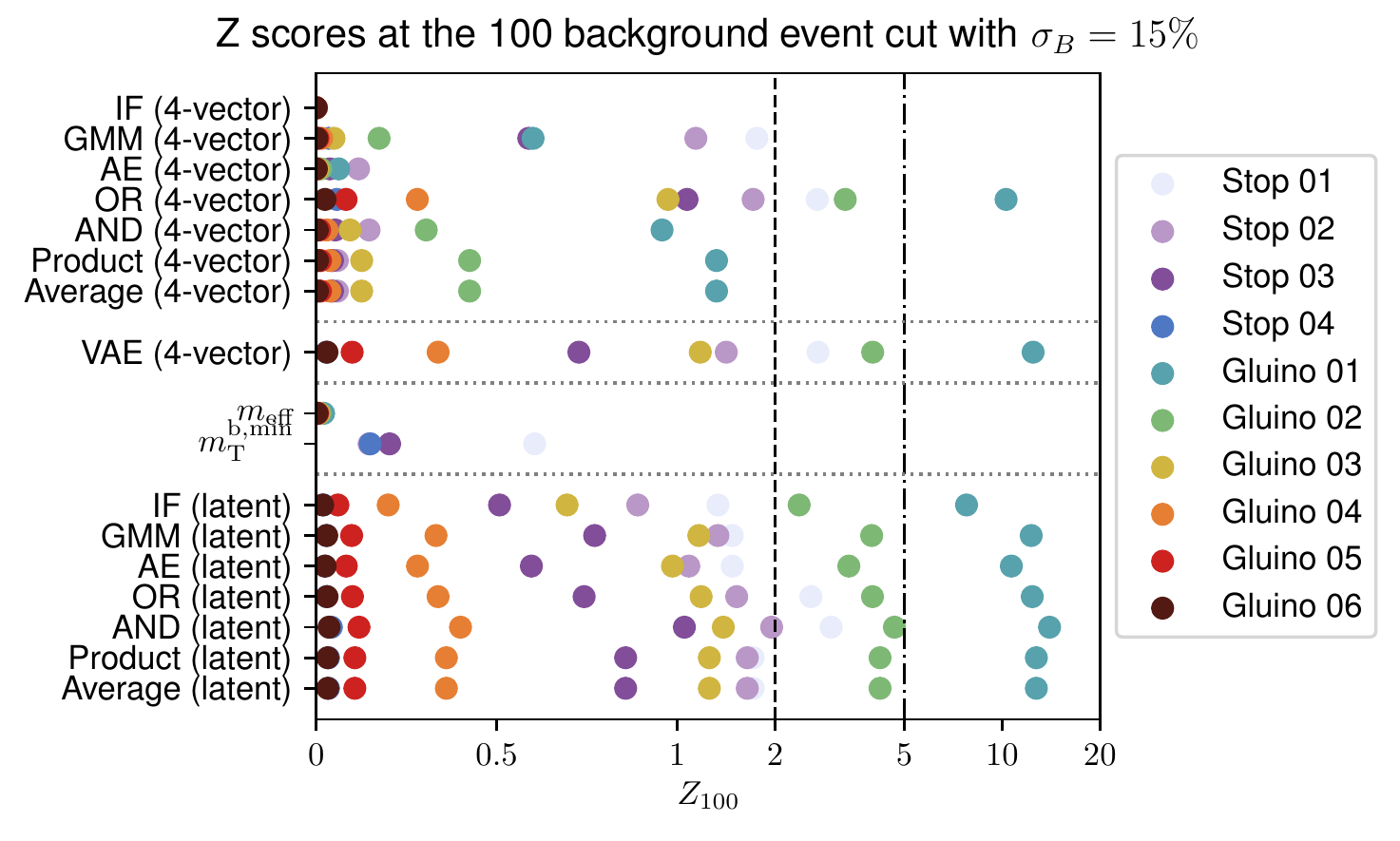}
 \caption{Z scores yielded from various algorithms applied to 4-vector components and latent space representations with a 15\% relative systematic uncertainty applied. See Table~\ref{tab:bsm-proc} for the signal definitions, and Table~\ref{tab:ascore_defn} for the definitions of the algorithms.}
 \label{fig:sigma15}
\end{figure}

So far, we have assessed the performance of each algorithm by considering their ROC curves. Figure~\ref{fig:sigma} displays side-by-side the significance values obtained using 4-vector components, and those obtained using the latent space variables. We can see that the performance of the IF, GMM, and AE increase when trained on the latent-space variables. Ultimately, training on latent space representations and performing an AND-combination of the normalised IF, GMM, AE and VAE anomaly scores yields the best performance for each considered signal of any technique detailed in this paper. Figure~\ref{fig:sigma15} displays the same significance values with a $15\%$ assumed systematic applied. The significances for all except the Gluino 01 signal are not high enough for discovery potential, although Stop 01, Stop 02, Gluino 01, and Gluino 02 are above the exclusion limit. This lack of discovery potential is not unexpected as these techniques do not optimise on the signal, and many of the higher mass signals are difficult to find even by conventional analyses. However, these results are quite promising for use as a preliminary step in a conventional analysis, especially since every anomaly score used in this analysis is minimally correlated with commonly used physical variables, however it is possible there are algorithms that yield anomaly scores that are correlated with these physical variables.

\clearpage
\section{Conclusions}
\label{sec:conclusion}
In this paper, we have examined a variety of unsupervised ML methods to perform anomaly detection in LHC searches. Our aim is to specify an anomaly score on an event-by-event basis that indicates how likely it is that the given event originates from new physics. We only address the problem of detecting \emph{anomalous} events, that is, events that are unlikely to have been produced by a SM process. The methods discussed in this paper are therefore not usable when the new physics shows up as an overproduction of a certain final state with similar kinematics to its respective SM background. 

The studied ML methods are the isolation forest (Section~\ref{subsec:IsoFor}), the Gaussian mixture model (Section~\ref{subsec:GMM}), the static autoencoder (Section~\ref{subsubsec:autoencoder}), and the variational autoencoder (Section~\ref{subsubsec:vae}). We have defined an anomaly score for each of these techniques (summarized in Table~\ref{tab:ascore_defn}), and assessed their performance in determining whether an event is anomalous by training them on the SM dataset published in Ref.~\cite{Brooijmans:2020yij}. We have then tested this against a collection of supersymmetric benchmark scenarios, summarized in Table~\ref{tab:bsm-proc}. The performance of each model is represented by the significance measure $Z_{100}$ (Eq.~\eqref{eq:significance}), which is the significance one obtains after cutting on the anomaly score that selects $100$ background events. 

In our training, we have first employed the 4-vectors of the events as inputs for the ML algorithms. Our results are summarized in Figure~\ref{fig:sigma} assuming a non-existent systematic uncertainty. The IF, GMM and the AE on their own show a rather poor separation across all models. The VAE is unique due to its clustering of the information contained in the 4-vectors in its latent space, which groups non-linear combinations of the 4-vectors. These non-linear combinations may be viewed as new observables, therefore, the IF, GMM and AE algorithms may also be trained on the latent-space variables. By doing this, their performance dramatically increases, however, their performance does not exceed that of the VAE itself. 

In addition to assessing the performance of individual ML methods, we also have explored combining these techniques in various ways. To this end, we have considered four different ways to combine their anomaly scores (Section~\ref{subsec:combinations}): AND, OR, product and averaging combinations. The performance of these combination depends on the signal and input representation (4-vector or latent-space variables). Using the 4-vector input representation, we find the best performance for the stop quark cases using the OR combination, while for the gluino events, the VAE gives the best result. When trained on the latent-space variables, this signal dependence drops out, and we find that the AND combination outperforms the other algorithms individually and the other combinations. The method that gives the best performance in the most signal-independent way is then:
\begin{itemize}
    \item Train a VAE on 4-vectors of SM background events.
    \item Train a selection of ML techniques (which does not have to be limited to the techniques discussed in this work) on the latent-space representations of the 4-vectors of the SM events.
    \item Normalise their anomaly scores $x_i$ and use $x^{\rm AND} = \min(x_i)$ to determine the anomaly score for a given event.
\end{itemize}

We have compared our results to using the physical variable $m_{\rm eff}$ ($m_{T}^{b,min}$), which is often used in gluino (stop) searches to discriminate background from signal events. The techniques outlined in this paper outperform the use of this observable. 

\section*{Acknowledgements}
MvB acknowledge support from the Dutch NWO-I program 156, "Higgs as Probe and Portal", and the Christine Mohrmann Stipendium. R. RdA acknowledges partial funding/support from the Elusives ITN (Marie Sk\l{}odowska-Curie grant agreement No 674896), and the Spanish MINECO grant ``SOM Sabor y origen de la Materia" (FPA 2017-85985-P). MW and AL are supported by the ARC Discovery Project DP180102209. 

\bibliographystyle{JHEP}
\bibliography{bibliography}

\end{document}